\documentclass[manuscript]{aastex}

\shorttitle{Photometric Study of IC 348}
\shortauthors{Cohen, Herbst \& Williams}

\begin{document}

\title{A Multi-Year Photometric Study of IC 348}

\author{Roger E. Cohen, William Herbst }
\and
\author{Eric C. Williams}
\affil{Astronomy Department, Wesleyan University, Middletown, CT 06459}

\begin{abstract}

The extremely young cluster IC 348 has been monitored photometrically over 5 observing seasons from Dec 1998 to March 2003 in Cousins {\emph{I}} with a 0.6 m telescope at Van Vleck Observatory. Twenty-eight periodic variables and 16 irregular variables have been identified.  The variability study is most sensitive for stars with I $<$ 14.3 mag; at that brightness level we find that 24 of the 27 known PMS cluster members in the monitored field are variables, illustrating the value of photometric monitoring for identifying PMS cluster members. Among this brighter sample, 14 of the 16 known K or M-type WTTS were found to be periodic variables, while all 5 of the known CTTS were found to be irregular variables. In the full sample, which includes 150 stars with I mag as faint as 18, we find that 40\% of the 63 WTTS are detected as variables, nearly all of them periodic, while 55\% of the 20 CTTS are also detected as variable, with {\it none of them periodic.} Our study suggests that 80-90\% of all WTTS in young clusters will be detected as periodic variables given sufficiently precise and extended monitoring, whereas CTTS will reveal themselves primarily or solely as irregular variables. This has clear consequences for PMS rotational studies based on photometric periods, suggesting that any such sample may be biased against stars which are currently actively accreting (i.e. CTTS).  We examine the stability of the periodic light curves from season to season. All periodic stars show modulations of their amplitude, mean brightness and light curve shape on time scales of less than 1 yr, presumably due to changes in spot configurations and/or physical characteristics. In no case, however, can we find definitive evidence of a change in period, indicating that differential rotation is probably much less on WTTS than it is on the Sun. While some stars show a hint of what could be cyclic behavior analogous to the sunspot cycle, no clear cycles could be found. It appears that most of the variation in light curve shape is caused by redistribution of spots on the surface rather than by an increase or decrease in the areal spot coverage. While most of the variables are of K or M spectral class, we do confirm the existence of three, low amplitude, periodic G stars. The rotation periods of these more massive stars are short compared to the bulk of the sample; it appears that mid-K to early M (i.e $\sim$0.5 M$_\odot$) represents a minimum in mean rotation rate for extremely young stars. Among the non-periodic stars, we report the detection of two possible UXors as well as a pre-main sequence star, HMW 15, which apparently undergoes an eclipse with a duration exceeding three years.

\end{abstract}


\section{Introduction}

It is well known that T Tauri stars (TTS) vary on many timescales, both periodicially and irregularly.  While a few photometric or spectroscopic observations can give us valuable ``snapshot'' information about the characteristics of these objects, more extended monitoring is needed to observe their full range of behavior and, ultimately, to understand it.  By analyzing changes in light curves over several years, we hope to gain an improved understanding of the physical mechanisms affecting the light of TTS.  While this endeavor requires large amounts of telescope time, CCD's have rendered such monitoring feasible with relatively small telescopes.  For over a decade, the 0.6m telescope at Van Vleck Observatory, located on the campus of Wesleyan University, has been used to monitor several nearby extremely young open clusters.  This consistent coverage facilitates a detailed study of a significant number of TTS over time. Here we present results based on five years of monitoring the nearby young open cluster IC 348.  

IC 348 is an ideal target for a variability study for several reasons.  It is both nearby and extremely young.  Its distance is 260$\pm$25 pc as calculated from {\emph{Hipparcos}} parallaxes \citep{s99} or 316 pc as determined by \citet{h98} using main sequence fitting.  The uncertainty in distance contributes to an uncertainty in age.  According to the models of \citet{dm}, the median age of the pre-main sequence stars in IC 348 is between 1.3 and 3 million years.  Photometric and spectroscopic observations of this cluster ranging across the electromagnetic spectrum have been amassed over the last decade. These include photometry and spectroscopy in the near-infrared by \citet{ll95} and \citet{lrll, l03}, and in the visual wavelength range by \citet{tj} and \citet{h98}.  A  wealth of X-ray data has been obtained from {\emph{ROSAT}} \citep{pz96} and {\emph{Chandra}} \citep{pz02}.  Membership probabilities have been determined by \citet{s99} from a proper motion survey, and a search for binaries using adaptive optics has been performed by \citet{dbs}.  

Results of the first six-month observing season of photometric monitoring of IC 348 at Van Vleck Observatory were presented by \citet{hmw}.  Here we present the results of four additional seasons of monitoring and discuss all of the data, with the aim of clarifying the nature of several different types of TTS variability over this timespan.  In addition to identifying new periodic and irregular variables, it is now possible to examine gradual changes in their light curves which may not be detected by observations which range over only one or two years.  Also, the determination of periods and phasing of light curves using observations from each season separately allows the first consistent observational investigation of the stability of TTS rotation periods and a search for spot cycles.   While \citet{s94} has claimed that TTS may exhibit differential rotation as dramatic as that of the sun, more recent studies, both observational \citep{ jk, b95} and theoretical \citep{kr97}, suggest that TTS rotate essentially as rigid bodies across their surfaces.  One of the goals of this study, made possible by its extended duration, is to empirically determine or constrain the degree to which TTS rotate differentially.  In addition, our data are useful for characterizing stars as WTTS and CTTS and for probing the distinctions between them.  In section 2 we discuss the observations and initial data reduction.  In section 3 we identify periodic variables and describe their characteristics.  Non-periodic variables, including the unusual eclipsing star HMW 15, are discussed in section 4.  In section 5, we summarize our findings and suggest some areas which would be useful for further study.   

\section{Observations and Data Reduction}

The observations were obtained between 10 December 1998 and 26 March 2003 with a 1024 $\times$ 1024 Photometrics CCD attached to the 0.6m telescope at Van Vleck Observatory, located on the campus of Wesleyan University.  Each pixel covers 0.6$\arcsec$ so our field of view is 10.2$\arcmin$ on a side.  On each clear night, a sequence of 5 one-minute exposures was taken through the Cousins {\em I} filter, as well as twilight flats, bias frames, and dark frames.  When possible, this sequence was repeated more than once per night.  Preliminary reductions were accomplished using standard IRAF tasks, and each set of five images was added together and shifted to the same position to within less than one pixel, creating one combined image with an effective exposure time of five minutes and an increased dynamic range.  A log of our observations is presented in Table 1.  The sample of 151 stars is the same identified by \citet{hmw}, but the coordinates given there were erroneous and have been corrected in Table 2.  

Before performing differential aperture photometry, all images with seeing worse than the chosen aperture radius of 6 pixels (3.6$\arcsec$) were rejected.  Seeing in the remaining images ranges from 1.5$\arcsec$ to 3.5$\arcsec$, with a median value of 2.5$\arcsec$.  Photometry was performed using the APPHOT package in IRAF, and the median level of the sky background was determined using an annulus with inner and outer radii of 10 and 15 pixels respectively.  There are some stars in our field whose photometry may be suspect because of their proximity to other stars, and a list of these objects can be found in \citet{hmw}.  They are also noted in Table 2.

Since our observations range over a five year period, it was desirable to determine a set of stable comparison stars which could be used over this entire timespan.  The comparison stars used by \citet{hmw}) did not fulfill this criterion, as they found that two of their comparison stars actually varied over small ranges.  To determine a single, consistent set of comparison stars for the entire observing interval, we began by finding a set of stable comparison stars for one season alone (1999-2000).  All stars which could not be measured photometrically on every night of observation were eliminated from the list of potential comparison stars, and an initial reference magnitude was defined by the remaining stars.  The remaining stars with the greatest variability were eliminated, and this process was repeated until a final list emerged with nine stars.  This procedure was repeated independently for a second season (2001-2002) and the two seasons had six comparison stars in common.  These six stars were subjected to our selection method for the remaining three seasons, and fortunately showed the same level of stability.  Using this method, which is not biased towards stellar characteristics such as spectral type or cluster membership, we emerged with stars 3, 7, 8, 10, 13, and 17 as our set of comparison stars for all five seasons.  While two of the comparison stars (13 and 17) show a slight trend in their magnitudes over the five-year course of the observations, their influence on the resulting differential magnitudes is negligible for two reasons.  First, the two stars change magnitudes in opposite directions, both remaining within a range of 0.05 mag.  Second, when averaged together with the other four comparison stars, deviations from season to season are diluted to a level of below 0.01 mag.  It is unclear whether these drifts are due to real variations of the stars or small changes in the wavelength dependence of the sensitivity function of the telescope, filter, and detection system over time.  

All six comparison stars display a standard deviation of under 0.01 magnitudes in each season, and their stability can be understood {\em a posteriori} as a result of their physical characteristics.  Five of the six stars are cluster members with spectral types earlier then G1, implying that they are on or close to the main sequence and can be expected to be relatively stable photometrically, and the sixth star (17) has been identified as a nonmember by \citet{s99}.  This set of comparison stars should, therefore, prove useful for future variability studies in this cluster.  

Differential magnitudes on the instrumental system ({\emph{i}}) were computed for each star on each night relative to the average magnitude of the comparison stars.  These were then transformed to a standard I magnitude ({\emph{I}}) adopting the color term derived by \citet{hmw}.  The adopted transformation is:
$${\emph{I}} = {\emph{i}} + 12.074 + 0.099 ({\emph{R}}-{\emph{I}})$$
Average {\emph{I}} magnitudes over the entire five years of observation are given in table 2, along with the R-I colors used for the transformation.  R-I values are taken from \citet{h98}, whose photometric system matches ours best, when available and from \citet{tj} otherwise.  For the three stars with known spectral types from the literature but no R-I colors (53, 69, and 88), R-I values were taken as an average of several stars in our sample with similar magnitudes and spectral types.  For the remaining 22 stars, R-I was assumed to be 2.00, which (while less than ideal) will suffice as an approximation since most of these stars are relatively faint ({\emph{I}}$>$15.0).  Stars with assumed, as opposed to measured, R-I values are indicated by parentheses in table 2.  The designations used here are cross-referenced with those used in recent studies by \citet{lrll, l03}. 

In addition to the average magnitude, we also give in table 2 the standard deviation ($\sigma$) for each star based on its photometric scatter.  These values allow us to identify non-periodic variable stars (see section 4), and such objects are signified by the label ``var'' in the column labeled ``Period''.  The time series data were also searched for periodicity, as will be described in section 3.  When significant periodicity was detected, the period is given in table 2.  In the case of stars for which slightly different periods were found in different observing seasons, which will be discussed in detail in section 3, the given period is an average over all five years of observation.  Note that periodic stars are easier to detect as variables than stars which vary irregularly because their variations are concentrated in a narrow band in frequency space.  Therefore, all stars are searched for periodicity, not just those whose $\sigma$ values mark them as variable.

\section{Periodic Stars}

\subsection{Determination of Periods and Properties of Periodic Stars}

To search for periodicity, the periodogram function was calculated for each star in our sample using the method of \citet{s82}, as formulated by \citet{hb}.  Since our data ranges over a timespan of 4.5 years, this function could be calculated for each observing season separately as well as for the entire duration of the observations.  We used the following criteria to assess whether a star should be regarded as periodic.  First, it must have a peak in its power spectrum with a power greater than 7.0, corresponding to a false alarm probability (FAP, as defined by \citet{hb}) below 2\%.  Second, it must possess one or both of the following attributes: i) a well-defined light curve characterized by small scatter and accompanied by a strong peak in the periodogram in at least one season, or ii) the highest peak in the periodogram (exceeding a power of 7.0) recurring in multiple seasons.  Based on these criteria, we have detected periods for 28 stars in our sample.  Nine of these represent new detections, and the remaining 19 were discovered by \citet{hmw}.  

\begin{figure}
\plotone{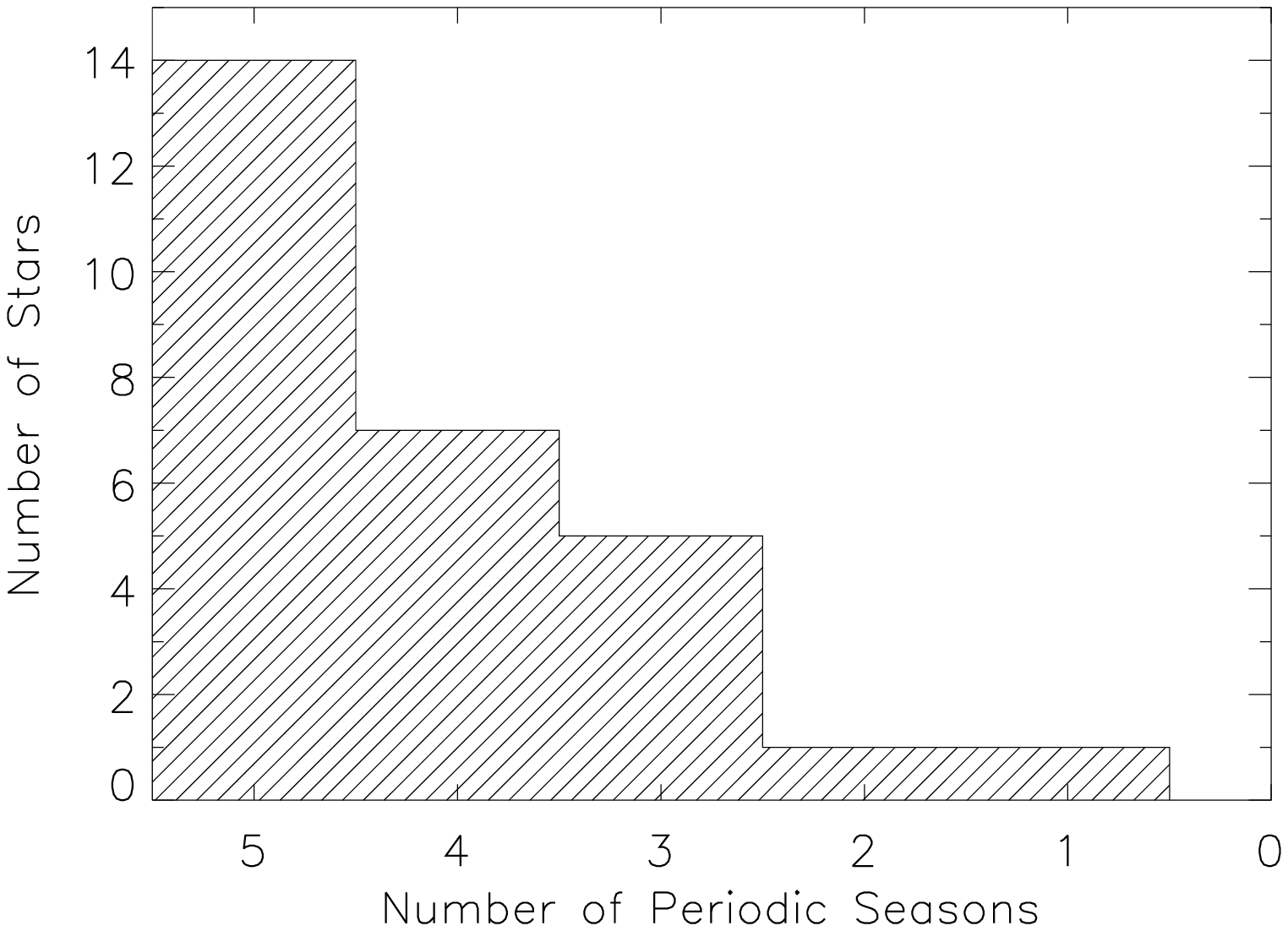}
\caption{A histogram showing how many stars were detected as periodic in each number of seasons.  Note that all periodic stars were periodic in at least two seasons, and 14 of the 27 periodic stars were periodic in all five seasons.}
\label{seashist}
\end{figure}

It is interesting to note that our sample is remarkably unambiguous with respect to periodicity.  All stars which were periodic in one season were periodic, with very similar periods, in at least one additional season except for one (star 47, which is discussed in section 4), and half of the periodic stars were so-identified in all five seasons.   Fig. \ref{seashist} shows a histogram of the number of seasons in which a star was detected as periodic.  A search was also performed for stars with smaller peaks in their power spectra, falling slightly below our criteria, but which might recur.  This was done by multiplying the power spectra from all five seasons at each frequency and searching for peaks in the result.  No additional periodic stars were found through this method, supporting our choice of selection criteria for periodicity.  It appears that periods of periodic variables in IC 348 are very stable, as is quantified below, and that our procedure for identifying them is effective and efficient.  

\begin{figure}
\plotone{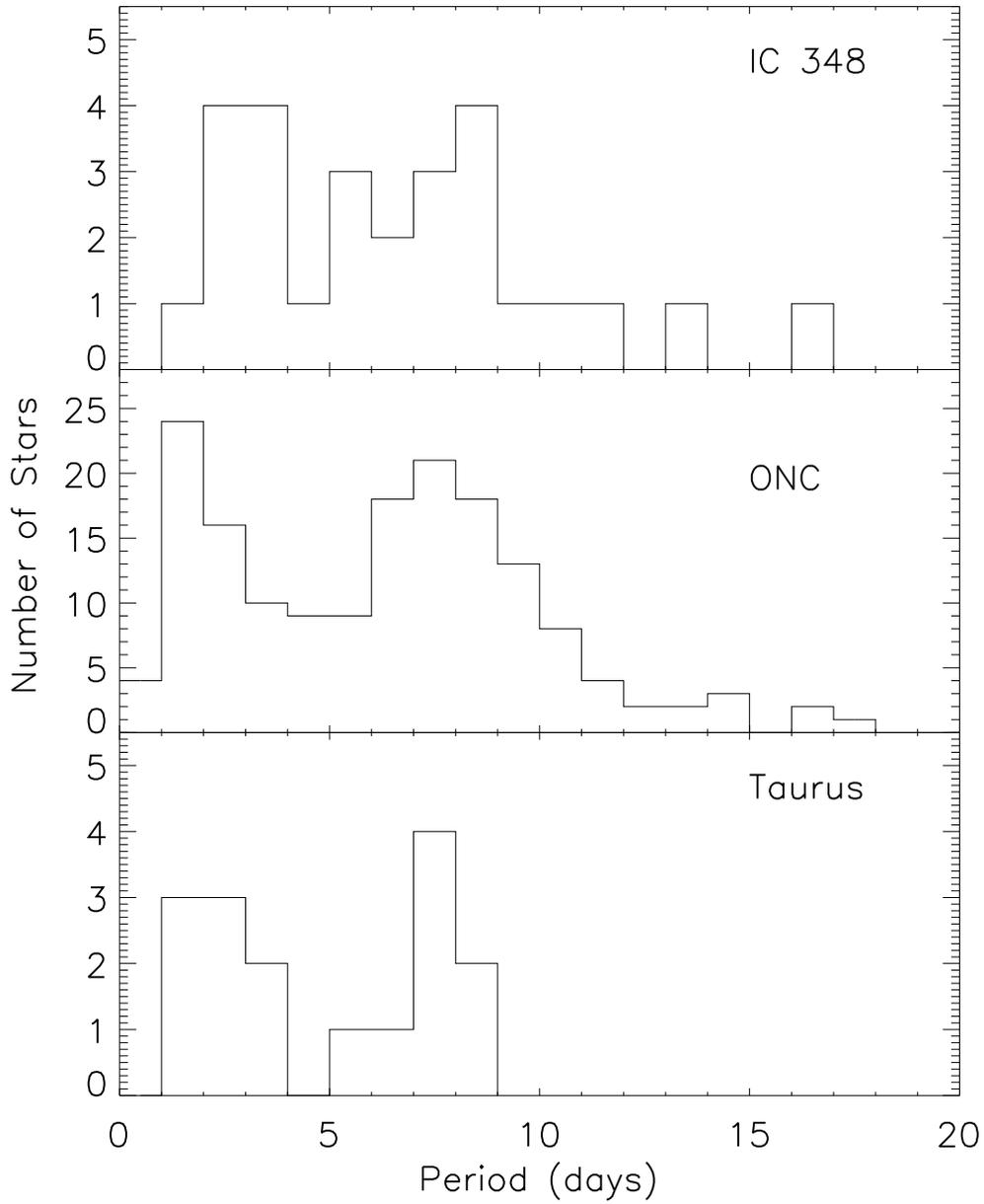}
\caption{Period distributions for our sample, the Orion Nebula Cluster, and the Taurus-Auriga association.  All three distributions show a paucity of stars with periods less than one day, between four and five days, or greater than about 10 days.  Only stars earlier than a spectral type of M2 are plotted.}
\label{perdisthist}
\end{figure}

The distribution of periods in our sample is shown in Fig. \ref{perdisthist} along with the period distributions of the Orion Nebula Cluster (ONC) from \citet{hbj} and Taurus \citep{b90}.  The distributions shown are, in all cases, for stars with masses above 0.25 M$_{\sun}$, a criterion fulfilled by all the stars in our sample except for one (star 42).  The significance of this division, which is equivalent to a spectral type of M2 according to the models of \citet{dm}, is that in the ONC stars below this mass have shorter periods and show a unimodal rather than a bimodal period distribution \citep{h00}.  While the number of periodic stars in IC 348 and Taurus-Auriga are too small to be clearly distinguishable as unimodal or bimodal, the IC 348 distribution shares three characteristics with the ONC and Taurus samples.  These are: i) an absence of periods shorter than one day, ii) a relative deficiency of periods between four and five days, and iii) a small number of periods greater than ten days.  More quantitatively, a two-sided Kolmogorov-Smirnov test \citep{ptvf} reveals that there is a 65\% chance that the period distributions of IC 348 and the ONC came from the same parent population, and a 53\% chance for IC 348 and Taurus.  Given the dramatic differences between the environments of the ONC, the Taurus-Auriga association, and IC 348, this similarity in period distributions has an interesting implication.  It appears that environmental factors such as stellar density or the presence of hot stars and ionized gas have little or no effect on the distribution of rotation periods for pre-main sequence stars with masses above 0.25 M$_{\sun}$.  

\begin{figure}
\plotone{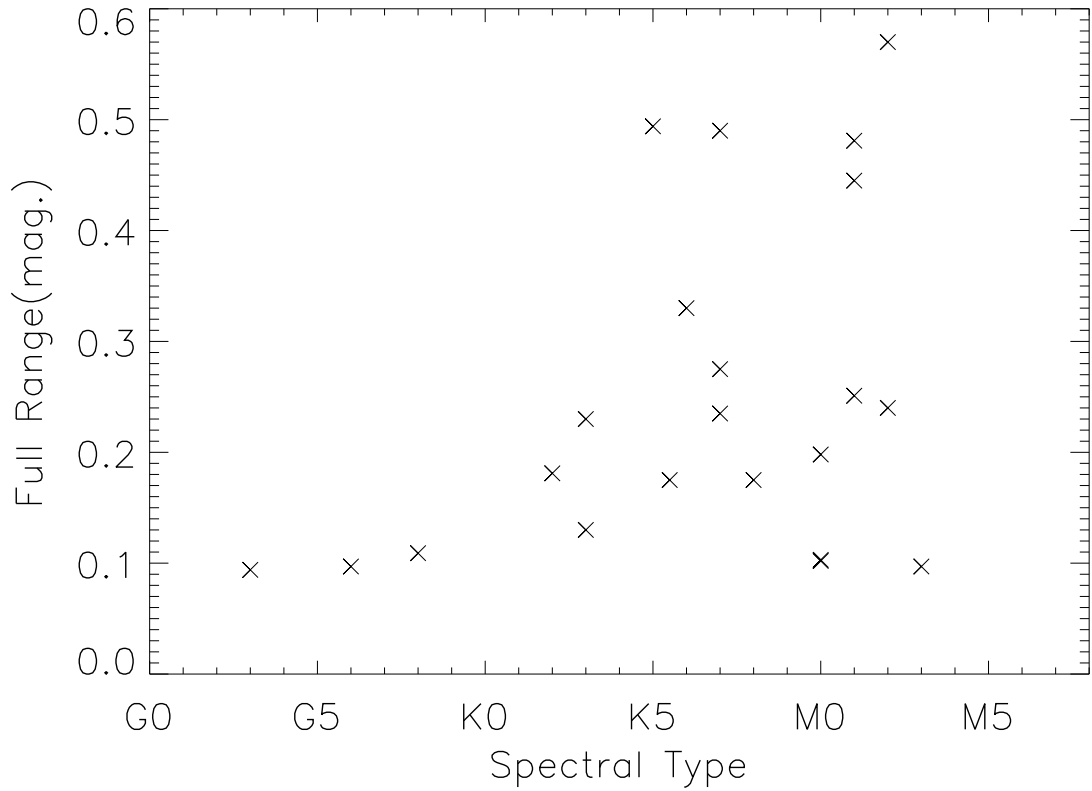}
\caption{The full range of variation over all five seasons vs. spectral type for periodic stars.  Stars subject to contaminated photometry are not shown, and star 47 is excluded since it displays large-amplitude irregular variability in the seasons in which it is not periodic (see section 4).}
\label{rangeVspt}
\end{figure}

In Fig. \ref{rangeVspt} we show the full range of variability for the periodic stars as a function of spectral type.  It is clear that most of the periodic stars have spectral classes from K to early M, which is typical of TTS in general.  The absence of later-type periodic stars is a selection effect.  They are too faint for us to obtain sufficiently reliable photometry at Wesleyan to find periods.  The relatively small number of G-type periodic variables and their small amplitudes is in basic agreement with what is found for PMS variability in general \citep{h94}.

\begin{figure}
\plotone{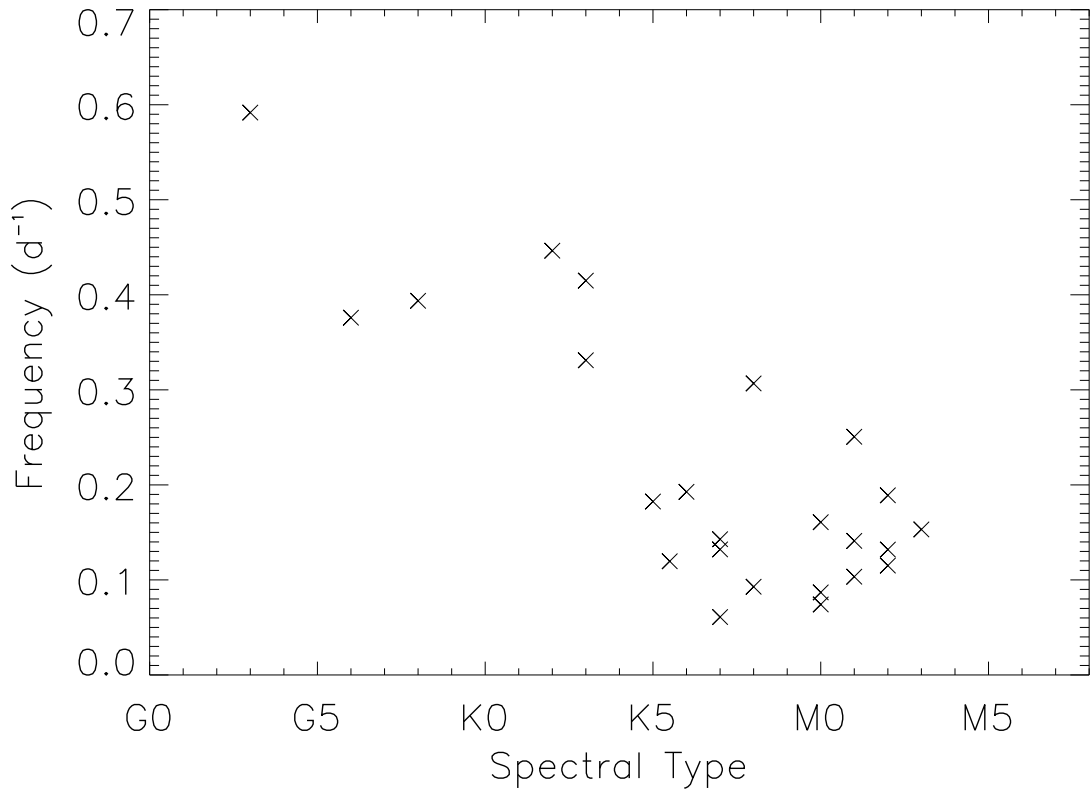}
\caption{Frequency vs. spectral type for the periodic stars in our sample.}
\label{perVspt}
\end{figure}

In Fig. \ref{perVspt} we show that there is an interesting correlation between rotation period (or frequency = 1/P) and spectral type in the sense that more massive stars rotate faster.  \citet{hmw} suggested a possible correlation between mass and rotation period, which is equivalent to a correlation between spectral type and rotation period.  While the sample may be too small to be definitive, it is nonetheless a striking correlation. A non-parametric (Spearman) rank-order test \citep{ptvf} indicates that spectral type is correlated with rotation in these data at a significance level of 99.7\%.  \citet{h01} showed that in the ONC, stars with spectral types later than M2 also rotate faster, as a group, than the mid-K to M2 stars. It appears, therefore, that mid-K and early M stars define a minimum in rotation rate among PMS stars of ONC and IC 348 age.  Both G stars and stars later than M2 (at least in the ONC) appear to spin faster, on the whole, although a wide range in spin rate certainly exists among PMS stars of all spectral classes.  Since disk-locking \citep{k91, os} is widely regarded as the braking mechanism for PMS stars, it appears that it is most effective in the mid-K to early M spectral regime, i.e. for masses $\sim$0.5 M$_\odot$. 

\subsection{Evolution of Light Curve Shapes}

Since our data extend over 4.5 years, it is possible to examine how the light curves of periodic variables evolve on that timescale.  Very little work has been done with regard to this question, a notable exception being the WTTS V410 Tau \citep{vhb, h89, g89, s03}.  One might expect both evolution of light curve shapes and changes in period, assuming TTS have some differential rotation.  Light curve shapes should evolve as the spots change size, shape, temperature, or location, and a latitudinal migration of spots could cause changes in period.  It would, of course, be very interesting to find any cyclic pattern which could be interpreted as evidence for a magnetic cycle.  For these reasons, we have carefully examined the data on our 28 periodic variables with respect to changes in period and light curve shape.
 
\begin{figure}
\plotone{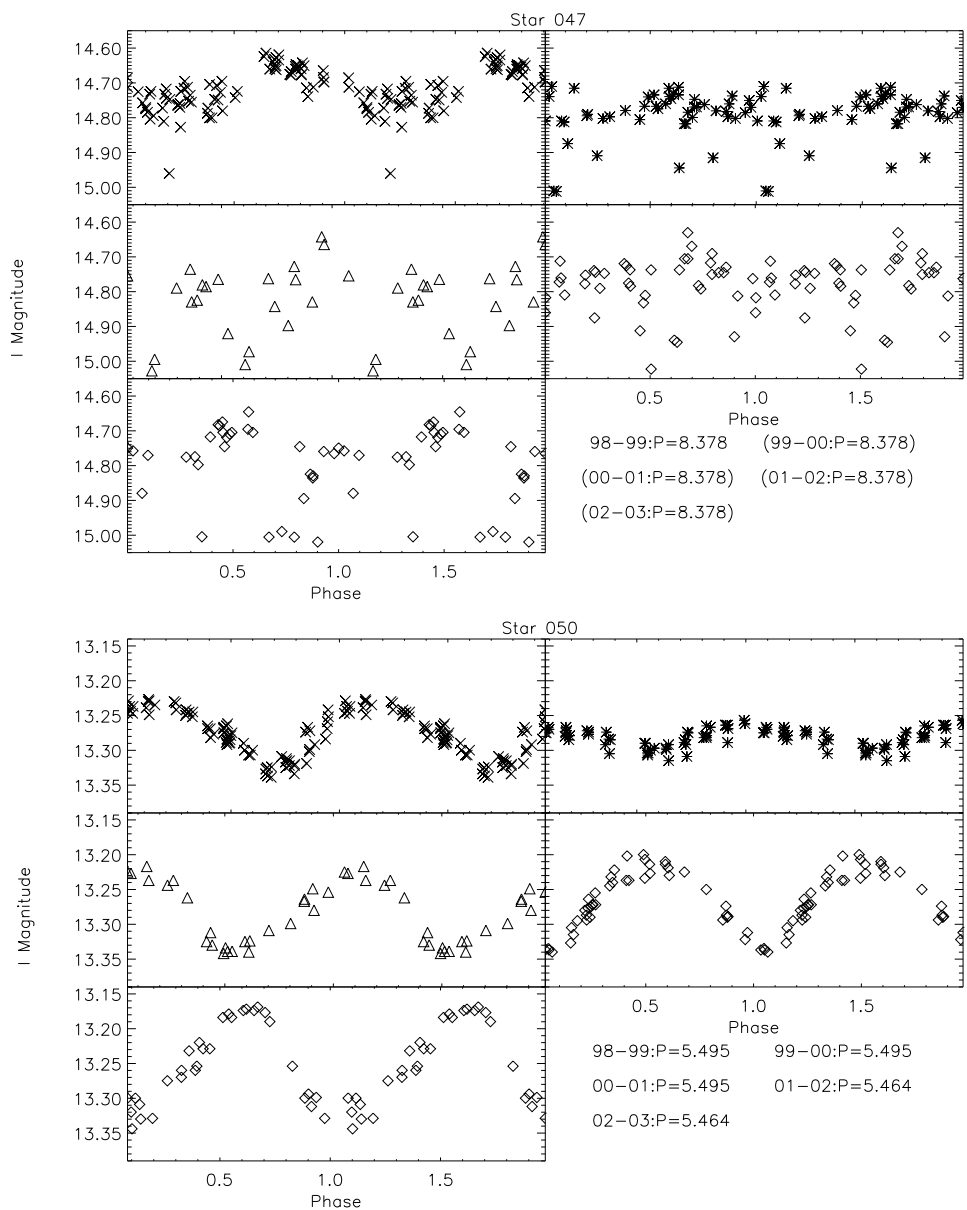}
\caption{Light curves for 10 of the periodic stars in our sample.  The period detected in each season is shown in the lower right of each plot.  Periods given in parenthesis indicate a season in which no period was found meeting our criteria for periodicity.}
\label{lcs}
\end{figure}

We began the analysis by searching for a single period which could be used to phase the entire data set for each star.  This was done by evaluating a single periodogram for each star based on all five observing seasons.  As noted above, we found that, in most cases, there was not a single strong peak at one frequency but several closely-spaced smaller peaks.  In no case could we find a period that produced a single convincing light curve when used to phase the data over the entire time span.  Evidently, the spots on all of these stars evolve on timescales of less than a few years, affecting either the period or the phase of minimum light or both.  In every case, we found that light curves phased using periods determined from individual seasons independently were clearly superior to any combined light curve spanning all five seasons.  For this reason, we present the data as individual light curves for each season, plotted with the best period (as indicated by the periodogram) for that season.  If a star did not have a peak in its periodogram meeting our criteria in a particular season, a period from another season was used to phase the light curve, and that period is given in parentheses in the figures.   Light curves for 10 of the 28 periodic stars in our sample are shown in Fig. \ref{lcs}.  We have chosen to show the most well defined or otherwise interesting light curves.  The full set of data may be obtained electronically by anonymous ftp to ftp.www.astro.wesleyan.edu. The files are in the directory bill/ic348/. 
  
Considering the most stable stars first, we note that 4 of the 28 periodic stars in our sample (16, 49, 69, and 134) had identical periods detected in all five seasons.  However, even these variables have light curves which show changes in shape, amplitude, and average magnitude between observing seasons.  Star 16 is the best example of stability in our sample (see Fig. \ref{lcs}), and even in its case there are obvious changes in the shape of the light curve and the phase of minimum light from season to season which no refinement in the period can eliminate.  Not a single star showed a precisely repeating light curve even from one season to the next, implying that the longest time that a given starspot configuration can remain stable enough to produce a completely coherent light curve is between $\sim$0.5-1.0 years.  For some stars, periods are detected in some seasons but not in others (e.g. star 12 in Fig. \ref{lcs}).  In these cases, the spot patterns on the stars evidently do not always remain stable for even the length of an observing season.  The stars most likely remain spotted, but the spot pattern must evolve such that a single period cannot be determined over six consecutive months of observation.  It is clear from these data that spots evolve in terms of size, shape or distribution on typical timescales of weeks to months ($\sim$few to tens of rotation periods), and we will now analyze the changes in starspot configuration in more detail.

\subsection{Spot Cycles?}

\begin{figure}
\plotone{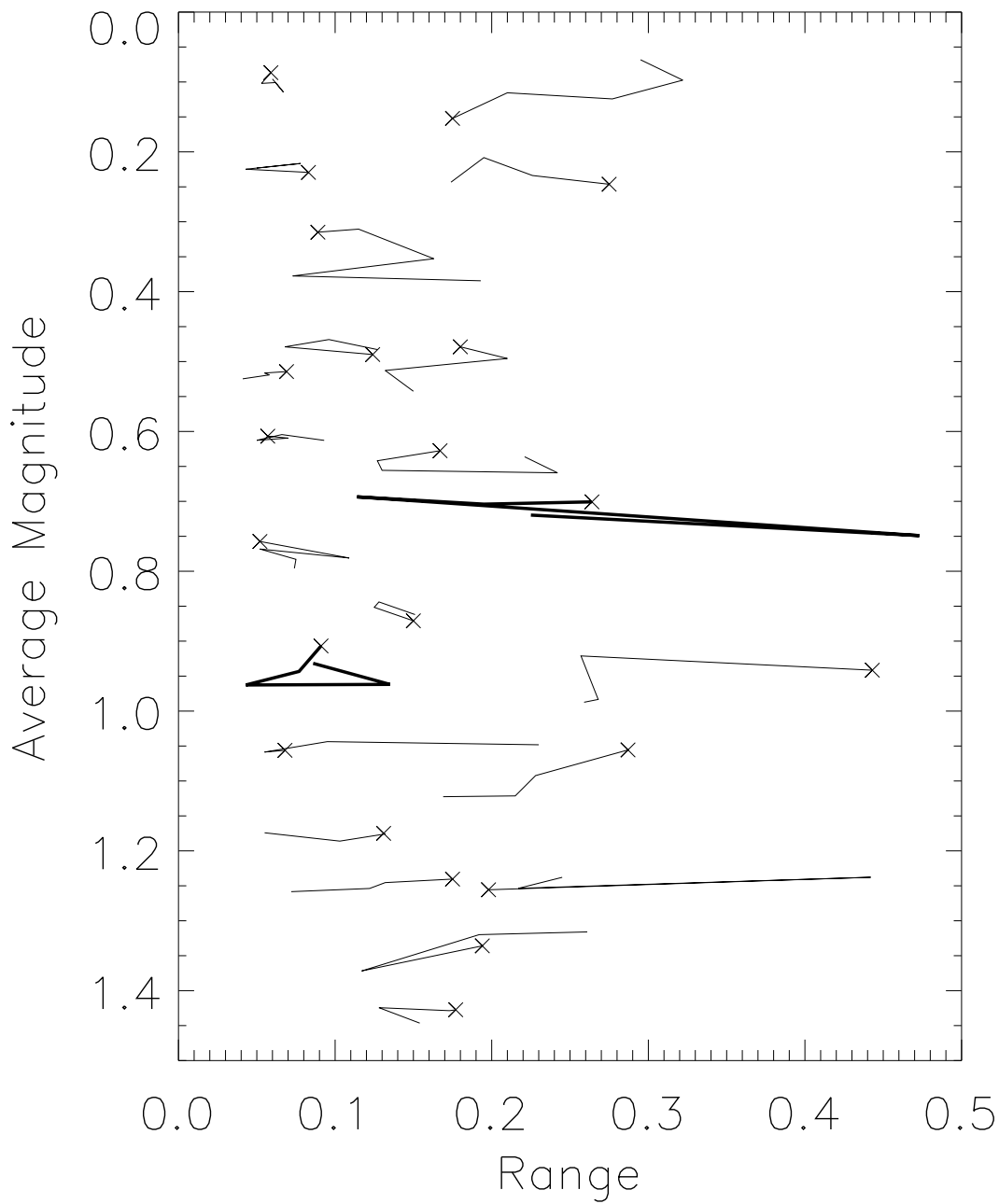}
\caption{Amplitude vs. average magnitude.  Each star is represented by a line which indicates its movement in the range-$\langle$I$\rangle$ plane over five seasons of observation.  The X marks the location of the first season for each star.  The magnitude scale is the same for each star, but their vertical location relative to each other is determined by clarity of presentation. Bold lines mark stars 30 (upper) and 12 (lower), which are discussed in the text. The range of star 30 would be much less if the faintest two points were eliminated (see Fig. 5). }
\label{spots}
\end{figure}

The changes in the light curves of the periodic stars in our sample can be accounted for by changes in starspot coverage and/or configuration, and one aspect of this phenomenon which remains relatively unexplored is cyclic behavior.  Since the sun displays cycles of magnetic activity on multiple timescales, the occurence or lack of occurence of such a phenomenon on pre-main sequence stars may shed some light on the physical mechanisms operating within these young solar analogs.  To investigate the extent to which spots on periodic TTS might exhibit cyclic behavior, the average magnitude was compared with the amplitude of each periodic star in each season in Fig. \ref{spots}.  The movement of stars in the range-$\langle$I$\rangle$ plane can be used to address the question of whether the total area covered by spots changes with time, as in the solar cycle, or whether spot redistribution is more important in changing amplitude.  In general, the paths of the periodic stars over five observing seasons on Fig. \ref{spots} are horizontal.  This stability of average magnitude from season to season most likely indicates that spots on TTS tend to redistribute themselves and undergo changes in size, temperature and location rather than appear or disappear {\emph{en masse}}. If the spot coverage were dramatically different from season to season we would expect the mean magnitude of the stars to fluctuate more than is observed. 

There are two stars, 12 and 30, which display potentially cyclic behavior based on their light curves and their movement in the range-$\langle$I$\rangle$ plane.  Their paths in Fig. \ref{spots} are shown in bold, and the possible cyclical behavior is also evident in their light curves (see Fig. \ref{lcs}).  In both cases, there is a gradual decrease in amplitude of variations, and in the case of star 12, there is no detectable period in the 2000-2001 season before its periodic variations return in the remaining two seasons.  Interestingly, when star 12 loses its periodicity, it remains stable at the fainter end of its magnitude range, not the brighter portion as would be expected based on a disappearance of cool spots.  It is possible, of course, that this could also be due to a redistribution of spots.  Star 30 exhibits the greatest change in range in our sample, although this is primarily caused by only two observations in the 2001-2002 season.  As with star 12, its range variations make sense when viewed in the context of its light curves.  However, having seen only one episode of spot disappearance and reappearance in each of these cases, neither constitutes solid evidence for a repeating cycle.  Although it is entirely possible that spot cycles on WTTS are either too long or too short to be detectable in our data set, our findings are consistent with those of \citet{kr97}, whose models of rotating T Tauri stars were not capable of producing the type of magnetic dynamo which would generate cyclical changes in spot configuration.  Since the sun exhibits spot cycles on multiple timescales, the shortest of which lasts approximately eleven years, continued monitoring will be useful to confirm and extend these results.  

\subsection{Stability of Periods}

As discussed previously, it was not possible to find a single period to coherently phase together all the data of any star in our sample.  In all but four cases, we found that different periods in different seasons were definitely required (in the other four cases one need only adjust the phase of the light curves, not the period).  On this basis we may claim to have detected real changes in periods for most of these stars.  However, as is now discussed, the size of these changes is very small and, in fact, on the edge of what can be considered significant.  We cannot rule out the possibility that changes in light curve shapes during a season cause most or all of the apparent changes in period detected.  The uncertainty in period for any season may be estimated using the method of \citet{k81}.  He relates the error in angular frequency, $\delta\omega$, to the amplitude ({\emph{A}}), the standard deviation of the noise ($\sigma_{N}$), the number of observations ({\emph{N}}) and the timespan over which the observations were made ({\emph{T}}) by:

$$ \delta\omega = \frac{3\pi\sigma_{N}}{2T(N)^{1/2}A} $$

\begin{figure}
\plotone{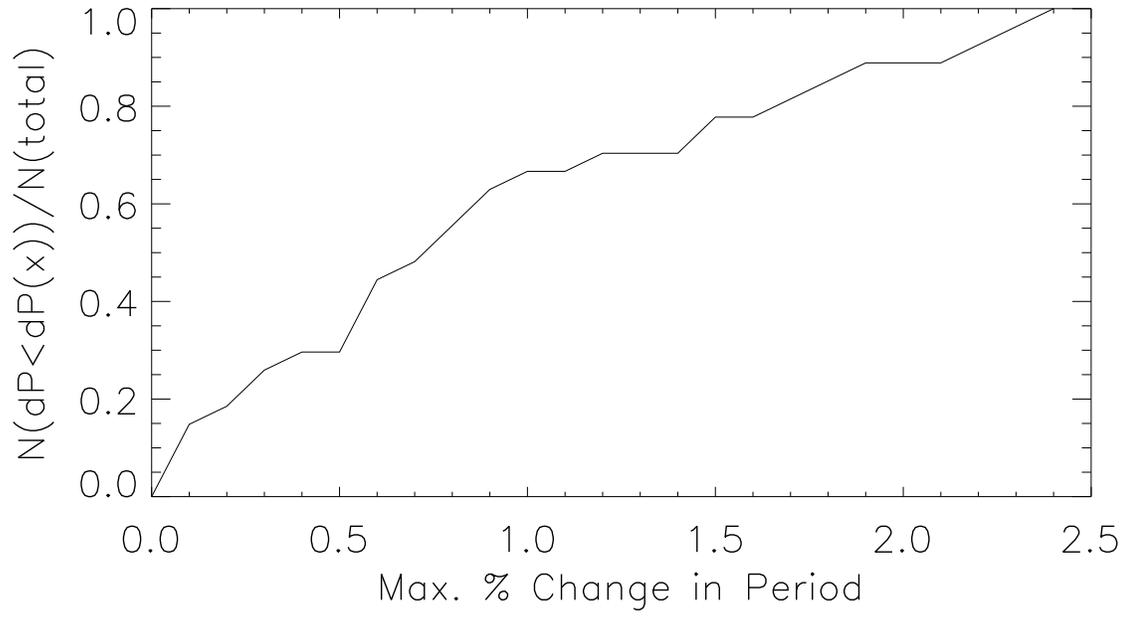}
\caption{A cumulative distribution of the period changes for the periodic stars in our sample.  The median is 0.8\% of the period.}
\label{dperdist}
\end{figure}

Application of this formula to our data suggests that our periods are accurate to between 0.0029 days and 0.26 days, or between 0.13\% and 1.58\%.  The changes in period seen between seasons range from 0 to 2.4\% and their cumulative distribution is shown in Fig. \ref{dperdist}.  While the two largest changes in period would be considered significant based on the error calculated using Kovacs' formula, we are not entirely convinced that we have detected any real period changes.  Examination of the light curves suggests that it is possible that phase changes are responsible for apparent changes in period.  

It is well known that, on the sun, the rotation period changes by $\sim$20\% from the equator to the poles.  The median observed period change for TTS in our sample is less than 0.8\%, suggesting that, unless the latitudes of the spots are confined to a very small range, TTS are significantly more rigid rotators than the sun.  This is in agreement with suggestions by \citet{kr97}, whose models of rotating TTS displayed differential rotation of under 1\%, as well as observational studies by \citet{jk} and \citet{rs96}.  Considering this evidence, it seems at least possible, if not likely, that during an observing season, spot configurations can change such that the period which best fits a series of observations shifts very slightly, causing a small shift in the peak seen in the power spectrum. We conclude that there is not yet any incontrovertible evidence for a period change in any TTS. On the contrary, periods appear to be remarkably stable on times of $\sim$5 yr. 

\section{Non-Periodic Variables}

Several classes of non-periodic PMS variability have been identified by various investigators over the years, including FUors, EXors, type II and type III (UXors) variations \citep{h77, he89, h94, hs98}. In some cases the time scales for these variations are quite long --- years or even decades. In addition, we might expect to find at least some examples of  eclipsing variables in a PMS sample, and it would be very important to identify them because of their potential to yield a great deal of otherwise unknowable information including basic stellar parameters such as mass and radius.  In this section we describe a search for non-periodic variables in our sample and discuss the properties of those discovered.  Most are ``normal'' (i.e. Type II) irregular variables, typical of CTTS.  A couple of possible UXors are also found but no FUor or EXor candidates. Two unusual variable stars (15 and 47) have been discovered which merit particular attention and are discussed separately in what follows.   

\subsection{Identifying Non-Periodic Variables}

\begin{figure}
\plotone{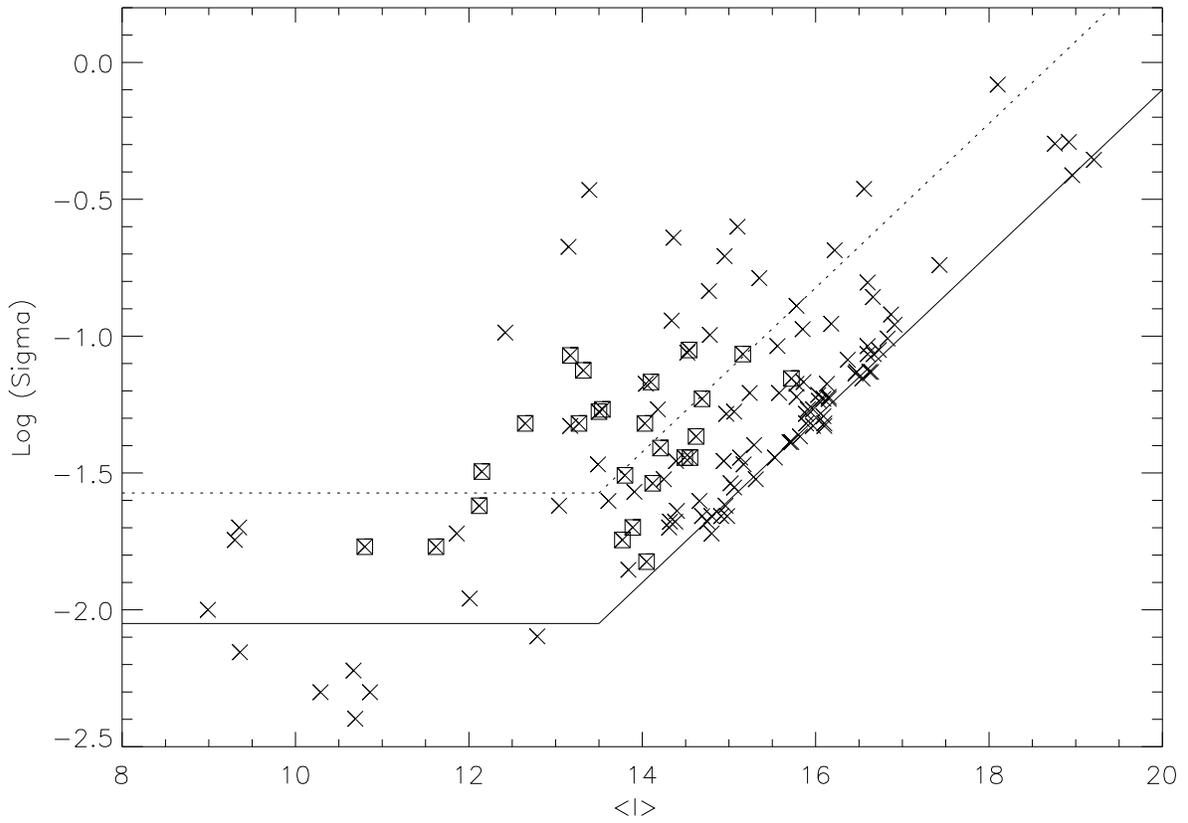}
\caption{Variability as a function of magnitude.  The solid line indicates the lower limit of $\sigma$ from \citet{hmw}, and the dotted line corresponds to 3$\sigma$.  Stars at risk of contaminated photometry are not shown, and boxed stars are periodic.}
\label{sigmas}
\end{figure}

 To characterize the degree of variability of the stars in our sample, we have used the standard deviation $\sigma_{var}$ measured over the entire timespan of our data.  In Fig. \ref{sigmas} we show log ($\sigma$) as a function of brightness.  Following \citet{hmw} we identify two regions on this diagram.  Stars brighter than {\emph{I}}$\sim$14 have a constant minimum error ($\sim$0.007 mag) dominated by systematic effects which are probably related to the flat-fielding process.  Fainter stars have an increasing minimum error due to the random effect of photon statistics.  The solid line represents a fit to the minimum errors taken from \citet{hmw}, which also describes our expanded data set well.  Four of the bright comparison stars fall below the minimum error line because their differential magnitudes are slightly correlated with their measured brightness on each night.   

To identify real variables we have adopted the criterion that a star must vary by more than three times the minimum $\sigma$.  This condition is shown on Fig. \ref{sigmas} as a dashed line.  We regard stars above the line as likely to be real variables.  Stars below it, especially fainter ones, may of course also be variables, but we do not have sufficient precision to distingiush them as such, unless they happen to be periodic variables. In that case, they will have already been detected as variables by the more sensitive periodogram technique discussed above. In all, there are 26 stars above the line, of which 10 are periodic variables,  leaving 16 ``irregular" variables, whose variations are discussed in this section. These stars are identified in Table 2 by a ``var" in the column labeled ``Period". A couple of light curves of special interest will be displayed here; all of the data are available through anonymous ftp from ftp.astro.wesleyan.edu, as noted above. 

A large majority (11/16) of the irregular variables identified in IC 348 by this study have the spectral characteristics of CTTS \citep{h98}. Only 2 are classified as WTTS by \citet{h98}, one is not yet classified and 2 are of G-type. One of the G-type irregular stars (18) is only barely above the line, but the other (15) is extraordinary and discussed in a separate section below. The two WTTS (56 and 101) clearly behave in somewhat anomalous fashion given their spectra; all of the other WTTS above the variability line are periodic variables. We have no explanation for the anomaly and simply note that these two stars may be worth looking at more closely in the future. Finally, one of the periodic TTS (47) shows large amplitude irregular variability as well. First we discuss the common type of variability (CTTS) and then the behavior of the more exotic stars 23 and 47. Finally, we briefly describe the amazing star 15, which has been the subject of a separate contribution \citep{chw}.

\subsection{Typical CTTS behavior: Type II variables}

Irregular variability was a defining characteristic of TTS (now CTTS) as originally described by Joy (1942,1945, 1947). Our study strongly supports the perhaps now canonical view that unsteady accretion, presumably from a circumstellar disk, drives these variations. Of the 13 known TTS in our field which could definitely be identified as irregular variables, 11 are CTTS. This is a clear confirmation of the correspondence which exists between an indicator of active accretion (H$\alpha$ equivalent width) and irregular variability. For the brighter end of our sample (I $<$ 14.3), where we can be reasonably certain that we have identified all of the variable stars present in the field, there are 5 CTTS  and every one of them is an irregular variable. The observed correspondence between the spectroscopic signature of a CTTS (equivalent width of H$\alpha$ $>$ 10 \AA) and irregular photometric variability is clear and compelling in this cluster. 

Infrared excess emission is also commonly used as an indicator of the presence of a disk and might also, therefore, be expected to correlate with irregular variability. Infrared photometry of IC 348 stars is given by \citet{ll95} in the {\emph{J}},{\emph{H}} and {\emph{K}} bands.  These data can be used in combination with the long-term photometric data which we have amassed to calculate a measure of infrared excess, following \citet{hi98}:

$$ \Delta(I-K) = (I-K) - (I-K)_{0} - 0.5A_{V} $$

The observed {\emph{I-K}} color is computed from our average {\emph{I}} magnitudes (taken over all five observing seasons) and the {\emph{K}} magnitudes of \citet{ll95}.  The intrinsic {\emph{I-K}} color is taken from \citet{kh95} based on the spectral types given by \citet{h98} or \citet{lrll}.  The extinction is calculated based on the {\emph{V-I}} color excess determined using the {\emph{V}} magnitudes of \citet{h98} or \citet{tj} otherwise.  Again, the calibrations of \citet{kh95} were used, and the extinction law adopted by \citet{hi97} was used.  Infrared excesses were calculated for all stars in our sample which had a spectral type determined by \citet{h98} or \citet{lrll}.  

\begin{figure}
\plotone{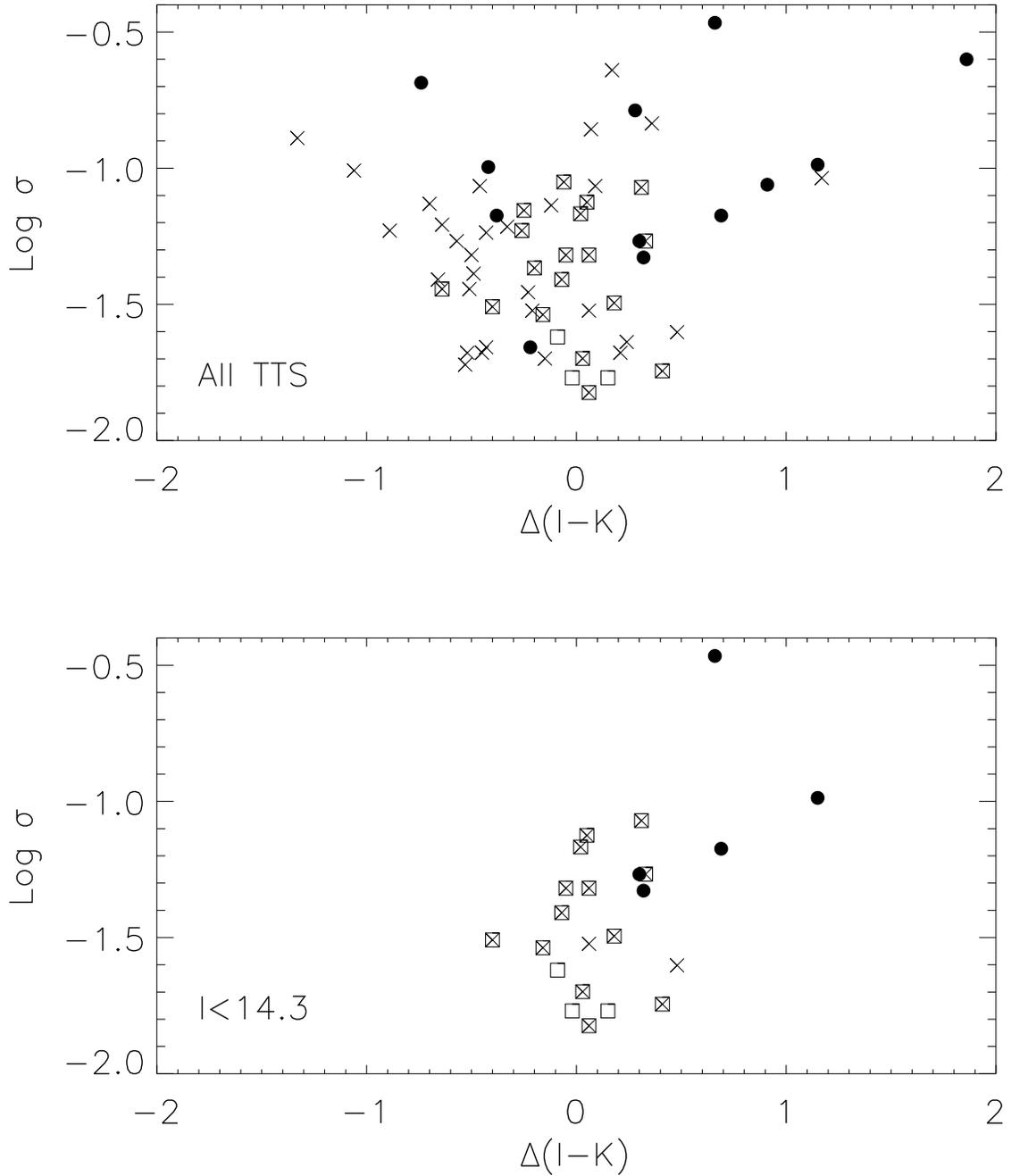}
\caption{Variability as a function of infrared excess.  Filled circles represent CTTS, crosses represent WTTS, and boxed stars are periodic.  The boxes which do not contain crosses represent the three periodic G stars in our sample.}
\label{SigIK}
\end{figure}

Fig. \ref{SigIK} shows variability, as characterized by $\sigma$, versus infrared excess, for all the TTS in our sample, as well as for only the TTS brighter than {\emph{I}}=14.3, where our identification of irregular variables may be regarded as complete.  The large scatter in these plots, including the non-physical phenomenon of infrared deficiencies represented by negative values of  $\Delta$(I-K), may result primarily from the non-simultaneity of the optical and infrared measurements.  Despite the scatter, additional causes of which will be mentioned below, some weak trends are visible in the full data set.  Of the six stars with $\Delta$(I-K)$>$0.5, five are CTTS, as one might have expected. Within the cluster as a whole, our data indicate that the CTTS are more variable than the WTTS.  These facts are consistent with the canonical view that CTTS variability is due at least partially to unsteady accretion from a circumstellar disk.  This becomes more evident when considering only the magnitude-limited sample.  In that case, all of the stars with $\Delta$(I-K)$>$0.5 are CTTS.  It is interesting to note that the periodic WTTS cluster around $\Delta$(I-K)=0.  They also appear to be somewhat less variable than the non-periodic WTTS and less likely to have derived infrared excesses below zero.  However, these latter two distinctions can be understood as selection effects due to the difficulty in detecting periods in fainter stars, as discussed above.  

\begin{figure}
\plotone{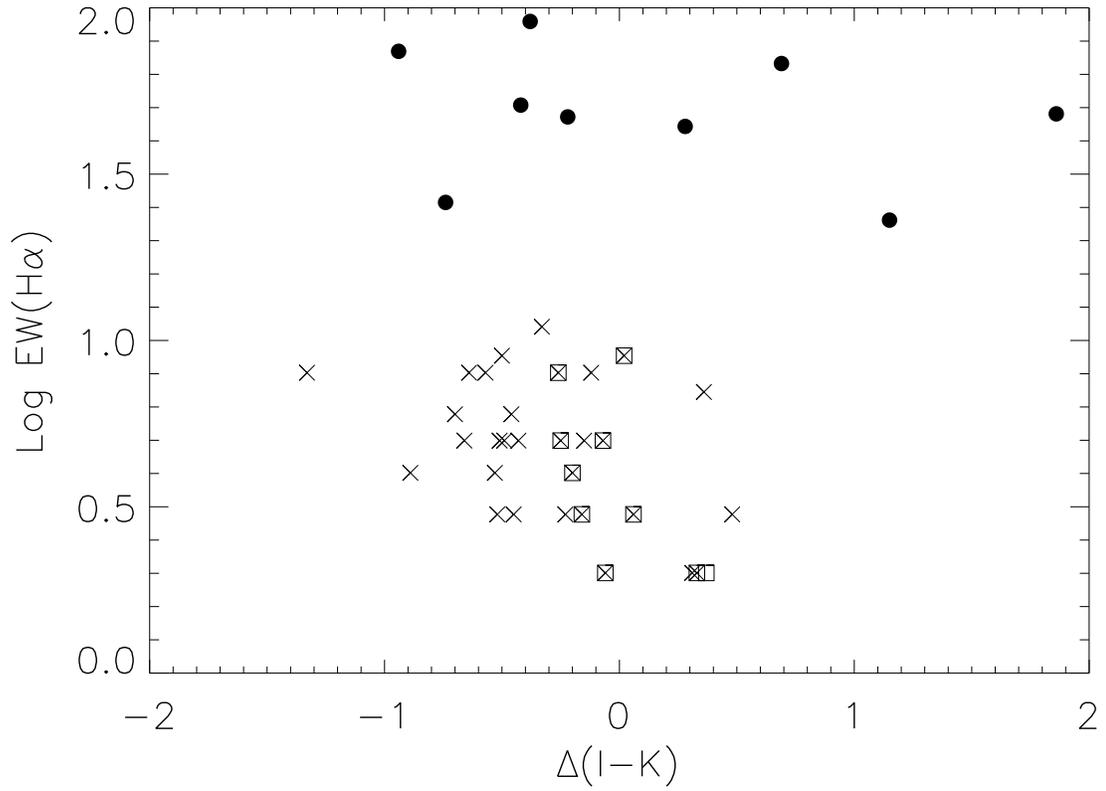}
\caption{Equivalent width of the H$\alpha$ line shown as a function of infrared excess.  Again, CTTS are indicated by filled circles and WTTS by crosses.  Boxed stars are periodic.}
\label{HAvIK}
\end{figure}

In general, the correspondence between infrared excess and variability is weak and stands in  contrast to the results for H$\alpha$ equivalent width described above. This led us to examine whether the two supposed disk indicators, H$\alpha$ equivalent width and infrared excess were in fact correlated with each other in this sample.  Fig. \ref{HAvIK} shows the log of the H$\alpha$ equivalent width, as measured by \citet{h98} versus the infrared excesses which we have computed for all the TTS in our sample with the neccessary spectral information.  While there is a fairly clear division between CTTS and WTTS, the correlation between infrared excess and hydrogen emission is  weak at best and actually almost non-existent.  Variability obviously correlates much better with H$\alpha$ equivalent width than infrared excess, perhaps because it is more a measure of current accretion activity or perhaps because it can be determined with better precision. In any event, in this sample, there is almost a one-to-one correspondence between irregular variability and  CTTS status based on H$\alpha$ equivalent width and only weak correspondence between infrared excess emission and irregular variability (or  H$\alpha$ equivalent width).    

It is also possible that the large degree of scatter in the infrared excesses shown in Figs. \ref{SigIK} and \ref{HAvIK} can be attributed to the causes discussed by \citet{hi98}, which include disk inclination effects, inner holes in disks, and excess K-band radiation from large cool spots.  However, there is another complicating factor in this case, namely the fact that the photometric measurements used to compute the infrared excess were not observed simultaneously or, in the case of K, very recently.  The infrared observations of \citet{ll95} were made in September 1991, and the photometric magnitudes of \citet{h98}, needed for the extinction computation, were calculated as an average of three observations between November 1993 and October 1996.  Based on the light curves of the CTTS in our sample it is clear that the photometric characteristics of a star can change from day to day and from year to year.  It is entirely possible that the general variability of the stars coupled with the non-simultaneity of the observations is a principle cause of the large scatter in the infrared excess measurement which then masks any correspondence with variability.  Only continued monitoring simultaneously in the optical and infrared could determine if this is the case.

\subsection{Some Unusual Stars}

\subsubsection{Star 23: an UXor?}

\begin{figure}
\plotone{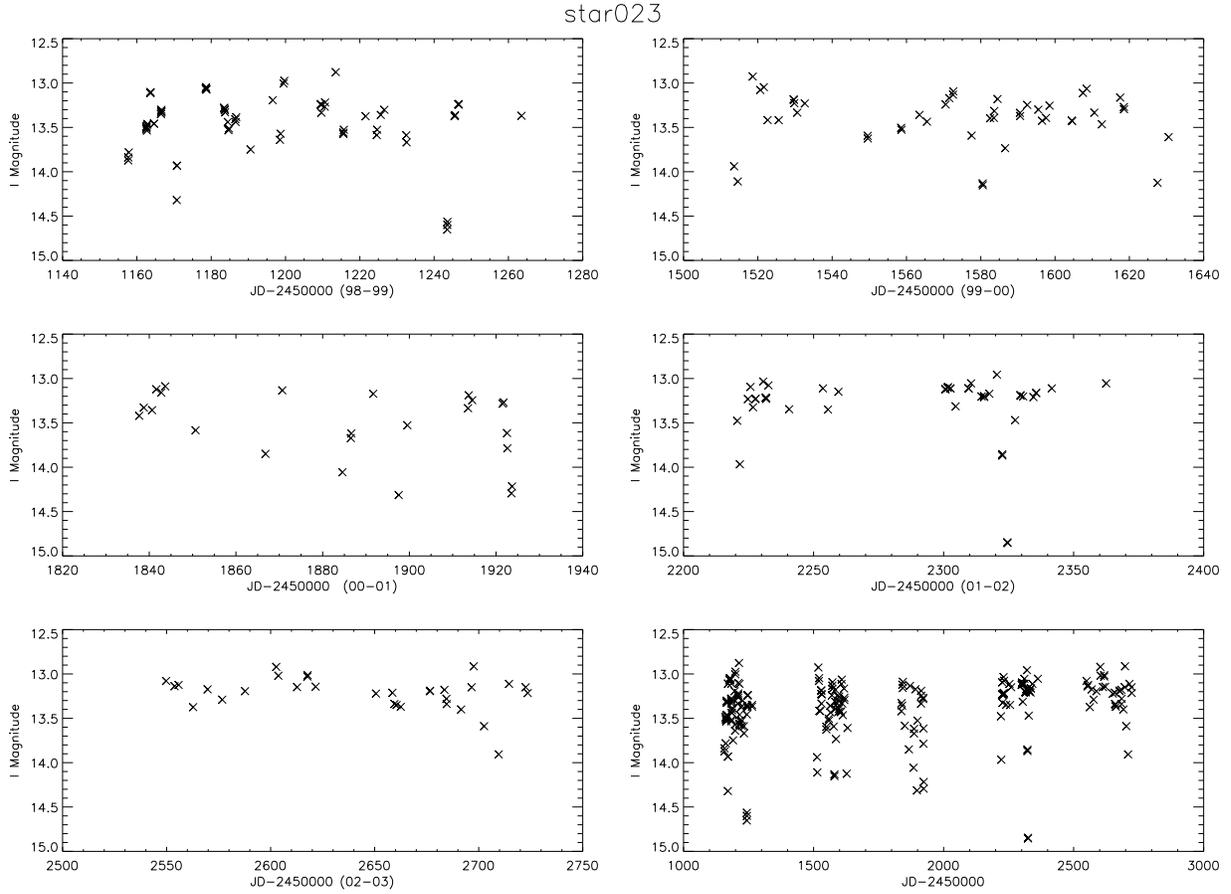}
\caption{The light curve of star 23.  Each season of data is plotted individually, and the bottom right plot shows all five seasons of data together on a compressed scale.}
\label{star23}
\end{figure}

This K6 CTTS was noted in Paper I as the most extreme variable star in our sample and that has continued to prove true. Its variations (see Fig. \ref{star23}), which approach 2 mag in amplitude, are also more reminiscent of UXor behavior than of typical CTTS behavior. In particular, the star tends to be near the bright end of its range most of the time and exhibits relatively brief, non-periodic excursions to fainter levels, as opposed to the more random sorts of fluctuations characteristic of Type II variables. On the other hand, most UXors are of earlier spectral type and do not have such strong H$\alpha$ emission. The star may be some sort of transition object or perhaps a close binary (or both). It clearly warrants further study as a potential UXor with interesting properties. Multi-color photometry and simultaneous spectral monitoring would be useful in elucidating its nature.   

\subsubsection{Star 47: Periodic and Irregular}

\begin{figure}
\plotone{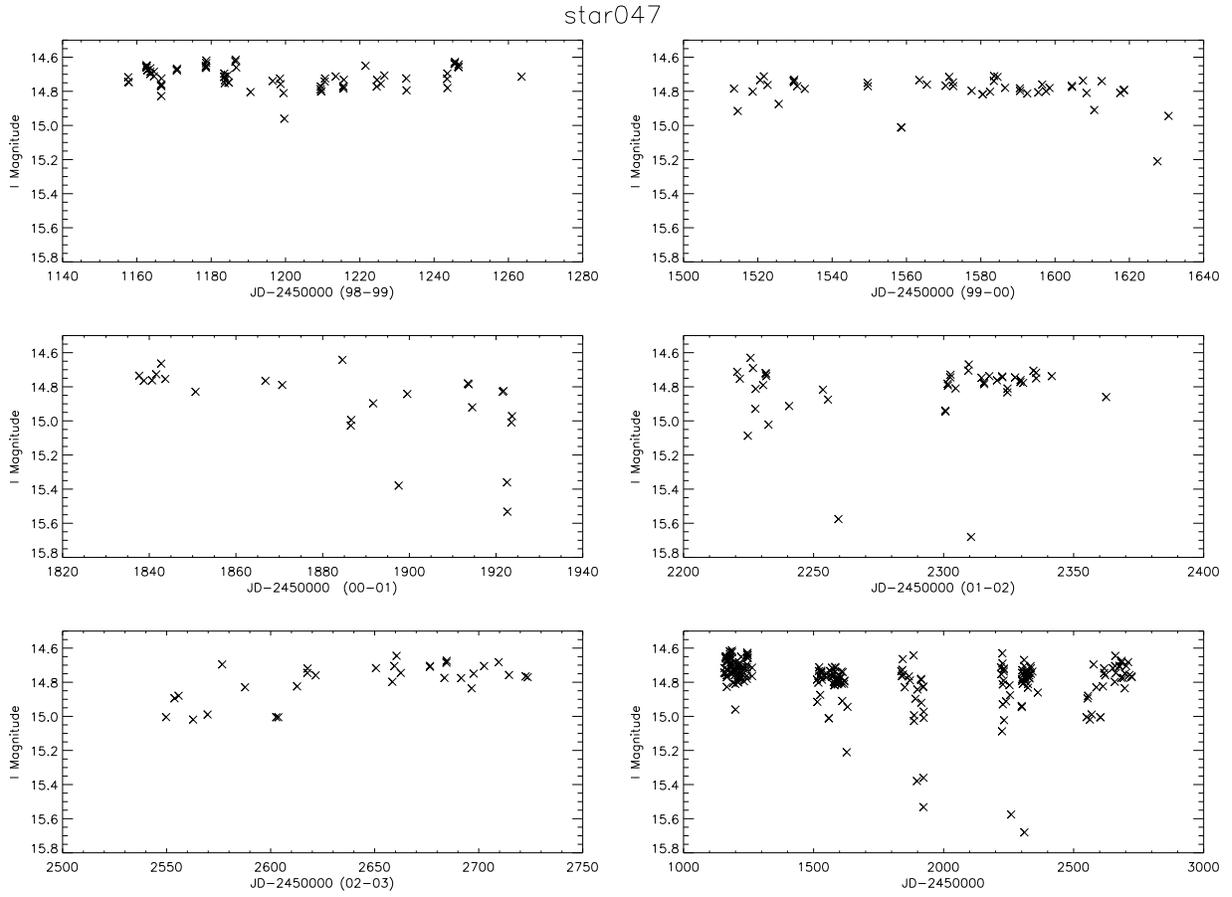}
\caption{The light curves of star 47 in each observing season.  This star is periodic in the 1998-99 season, and varies by less than 0.5 mag except for the four minima at JD 2451897, 2451922, 2452259 and 2452310.}
\label{star47}
\end{figure}

Star 47, a K8 WTTS, is the only example of a star in our sample which could be classified as both a periodic and an irregular variable (of substantial amplitude) at different times.  It was periodic in the 1998-99 season with a period of 8.38 days (see Fig. \ref{lcs}), but in the following seasons there are no peaks in its power spectra meeting our criteria for periodicity.  Between 2000 and 2002, its light curve is punctuated by several dramatic minima of increasing depth, the deepest occuring at JD 2452310 (see Fig. \ref{star47}).  The star fades by about one magnitude in one day, and four days later it has returned to its normal brightness. As in the case of star 23, this kind of variation is much more characteristic of UXor behavior than of typical CTTS (or WTTS) variability.  Again, it may be some sort of transition object or a binary and clearly warrants follow-up study. It is possible that this star, and star 23, are additional (to star 15, see below) examples of objects whose variability is caused by transient occultation events from dust clouds along the line of sight (and probably circumstellar or circumbinary). 

\subsubsection{Star 15: a Three-Year Eclipse?}

Within our sample, star 15 is the only example of a star which exhibits coherent behavior over 4.5 years of observation on a time scale longer than a single observing season.  This remarkable object apparently undergoes an eclipse lasting $\sim$3.5 years, the longest known astronomical eclipse ever detected, in which it gradually fades by 0.66 mag in {\emph{I}} and subsequently recovers to its out-of-eclipse magnitude.  Such an amazing and unexpected behavior clearly deserves special attention in the literature and was given it by \citet{chw}, to which the reader is referred for more detail.  Here we present a brief summary of that work and some supplementary material. Table \ref{table3} gives our photometry for the star.

\begin{figure}
\plotone{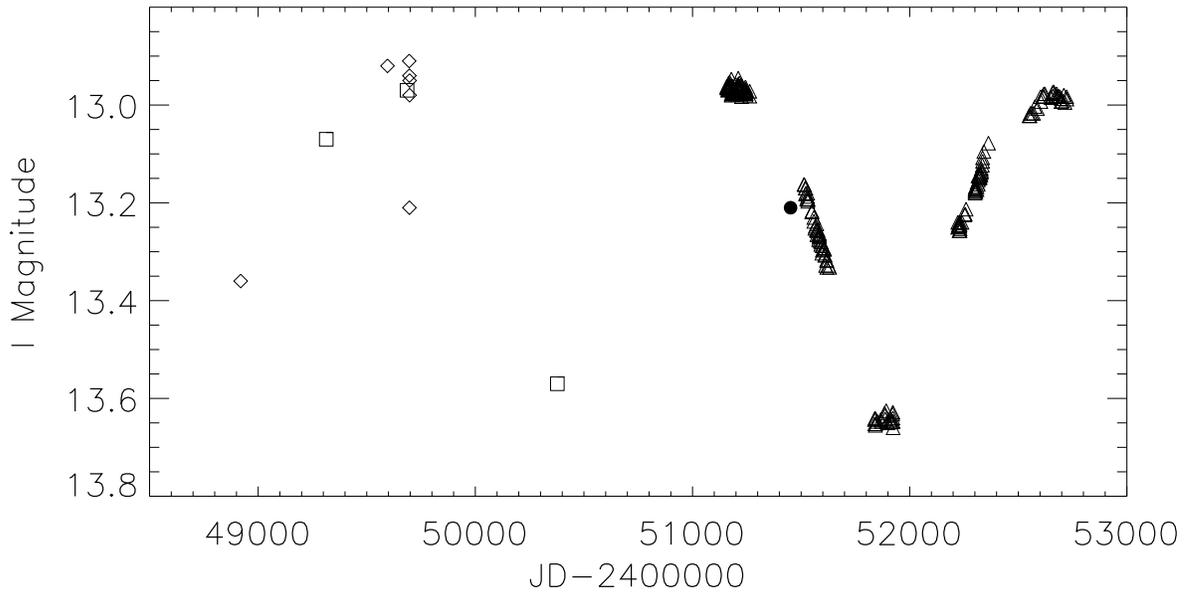}
\caption{The eclipse of star 15, shown with previous observations.  Our data are shown as triangles, previous observations by \citet{h98} as squares, and observations by \citet{tj} as diamonds.  The small filled circle represents an observation by \citet{l03} during ingress.  While most of the data are consistent with our out-of-eclipse magnitude, it is possible that this may be a periodic event with a period as short as four years.}
\label{15alldata}
\end{figure}

\citet{l03} have classified star 15 as G8-K6 based on near-infrared spectroscopy. It lacks infrared excess and hydrogen emission but is almost certainly a cluster member based on its locations on the sky and CM diagram.  The light curve of the star, based on our data and some from the literature, is shown in Fig. \ref{15alldata}.  The shape, consisting of a gradual ingress, a stable minimum, and a gradual egress, resembles that of an eclipsing binary with the obvious exception of its duration.  Observations in the {\emph{R}} band during minimum light show that the system became redder as it faded, although not as much as would be expected based on in interstellar reddening law. \citet{chw} argue that the star may be a binary system in which one member is occulted by a dust cloud, presumably of circumstellar or circumbinary origin. As Fig. \ref{15alldata} suggests, the eclipse may be recurrent on a time scale as short as 4 years. Obviously this is a star which should continue to be monitored photometrically and high resolution spectroscopy obtained near a brightness maximum should reveal whether it is, in fact, a binary system.

\section{Summary and Discussion}

Five years of monitoring stars in IC 348 has yielded a set of 28 periodic variables, almost all of which are WTTS, and 16 irregular variables, the large majority of which are CTTS. Our sample of variables is likely to be complete for stars brighter than I = 14.3. Of the 42 stars fulfilling that criteria (which includes 25 variable stars), 10 are of earlier spectral class than G and none of those 10 was detected as a variable. One is a late-type non-member and it was also not detected as a variable. Four of the remaining stars have no spectral information: one of these is a periodic variable and three are non-variable. That leaves 27 G, K or M-type PMS members of the cluster within our field which are brighter than I=14.3, and 24 of them (i.e. $\sim$90\%) were detected as variable stars in this study. We believe this nicely illustrates the power and value of variability work for identifying PMS stars in young clusters. 

It is furthermore interesting that, where our sample is complete (I $<$ 14.3), every one of the 5 CTTS was identified as an irregular variable, and 14 of the 16 K and M-type WTTS were identified as periodic variables. In addition, 3 out of the 6 bright G-type PMS stars showed periodic variations and 2 of the remaining 3 showed irregular variations (including the unusual variable star 15). Not only are variability studies excellent at identifying PMS stars, they are also apparently capable of distinguishing CTTS from WTTS with rather good precision. All of these results are quite similar, both qualitatively and quantitatively to what is found for the set of bright TTS in Tau/Aur, Orion, Sco/Cen and other well known associations analyzed by \citet{h94}.

One particular fact worth noting is that not one of the 20 known CTTS in our monitored sample revealed periodic variability during five seasons of monitoring. By contrast, 23 of the 64 known WTTS in the same area were detected as periodic and we probably would have found many more periodic variables among them if we had deeper images. This means that rotation studies based on periods determined by photometric monitoring are probably strongly biased against stars which are currently actively accreting. Since disk-locking is thought to be a primary controlling factor in the rotational evolution of PMS stars, and since actively accreting stars (CTTS) are, one might suppose, the ones most likely to be currently locked to their disks, this bias is important to consider when analyzing rotation data on clusters. In principle, one could use v sin i measurements to ascertain whether the rotational properties of CTTS differ from WTTS. \citet{rhm} found no significant difference between their periodic and non-periodic stars samples in the Orion Nebula Cluster, but they did not examine separately the definite irregular variables. A v sin i study of IC 348 could be extremely useful in addressing this issue. 
 
\acknowledgments

We thank G. Herbig and C. Jordi for providing us with unpublished data on HMW 15. We also thank the large number of Wesleyan students who contributed to this project over the years by obtaining the data on our campus telescope. This investigation was supported by grants from the NASA-Origins program.

\clearpage

\clearpage

\clearpage

\clearpage

\begin{figure}
\plotone{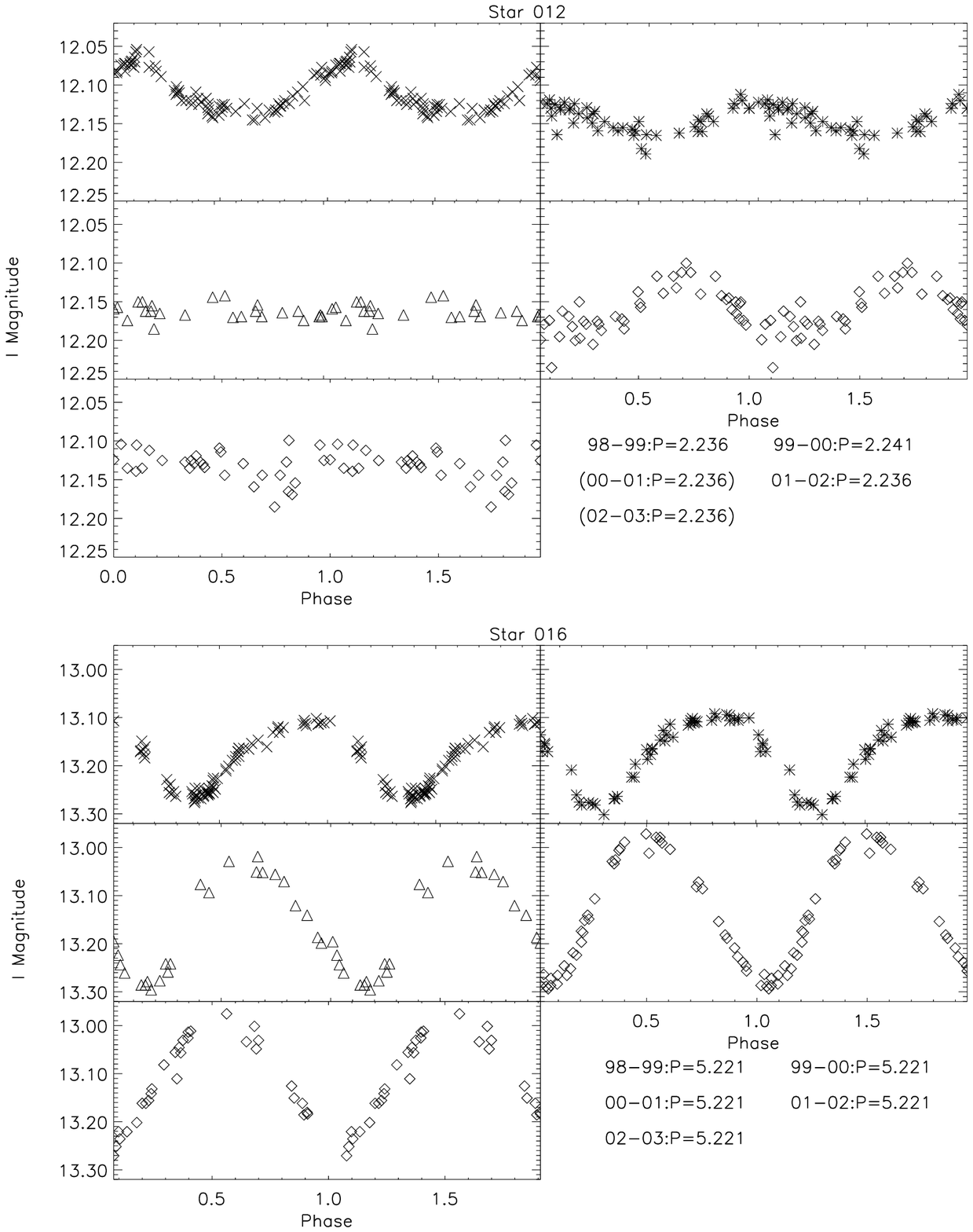}
\end{figure}
\begin{figure}
\plotone{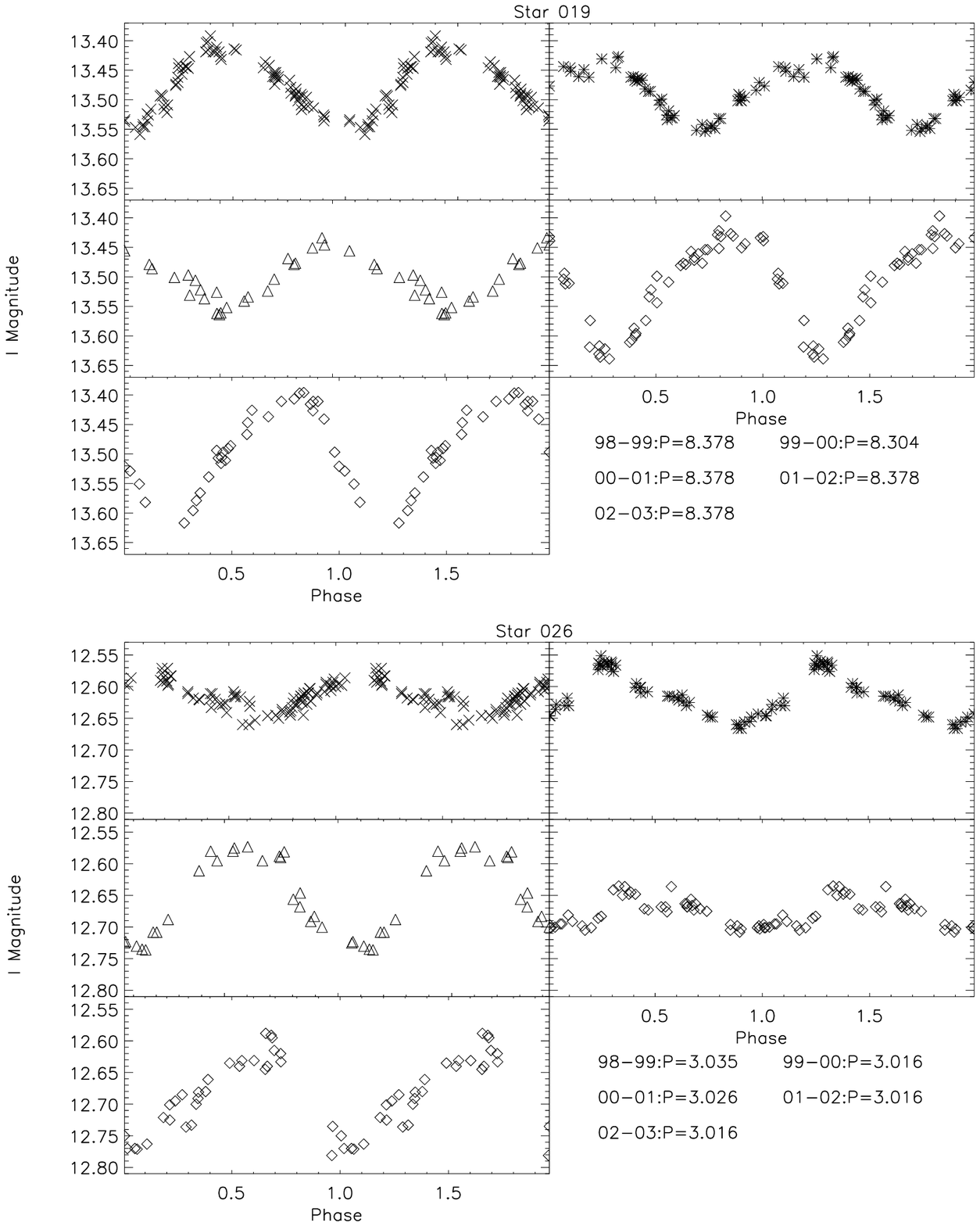}
\end{figure}
\begin{figure}
\plotone{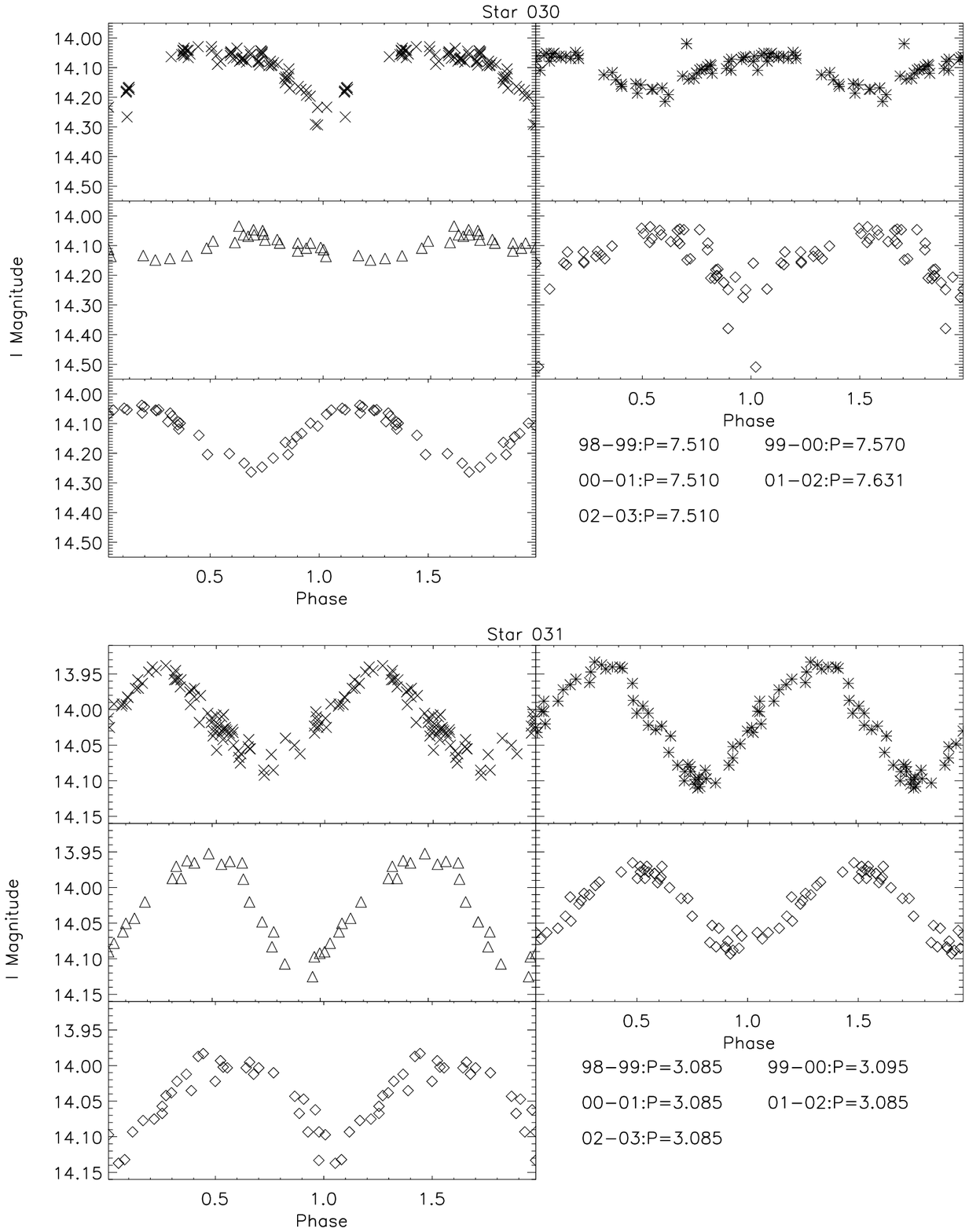}
\end{figure}
\begin{figure}
\plotone{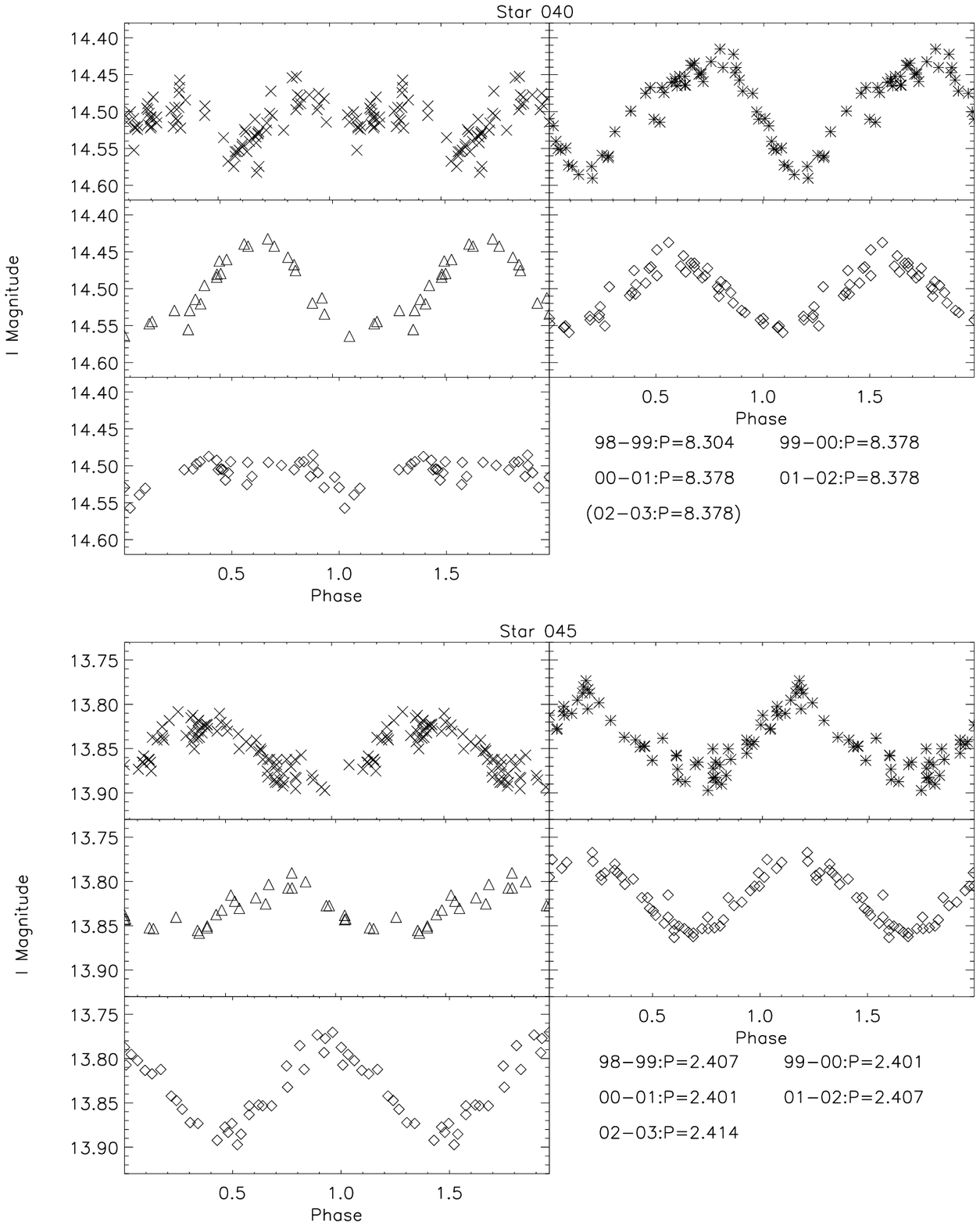}
\end{figure}
\clearpage

\clearpage

\clearpage

\clearpage

\clearpage

\clearpage

\clearpage

\clearpage

\clearpage

\begin{deluxetable}{l|llllllll}
\tablenum{1}
\tabletypesize{\scriptsize}
\tablecolumns{9}
\tablewidth{0pt}
\tablecaption{Observations.\label{tab1}}
\tablehead{
\colhead{\bfseries{Season:}}&&&&\colhead{\bfseries{JD-2450000.000}}\\
}
\startdata
\bfseries{1998-1999}& 1157.687 & 1157.695 & 1157.824 & 1162.581 \ 1162.589 & 1162.603 & 1162.611 & 1162.680 \\
& 1162.805 & 1162.813 & 1163.636 & 1163.644 \ 1163.658 & 1163.666 & 1164.598 & 1164.607 \\
& 1166.505 & 1166.513 & 1166.581 & 1166.589 \ 1166.610 & 1166.618 & 1170.748 & 1170.790 \\
& 1170.799 & 1178.561 & 1178.569 & 1178.603 \ 1178.611 & 1178.702 & 1178.710 & 1183.496 \\
& 1183.522 & 1183.627 & 1183.669 & 1183.727 \ 1184.464 & 1184.657 & 1184.782 & 1186.525 \\
& 1186.599 & 1186.745 & 1190.588 & 1196.564 \ 1198.515 & 1198.683 & 1199.484 & 1199.690 \\
& 1209.515 & 1209.534 & 1209.592 & 1209.686 \ 1210.569 & 1210.649 & 1213.526 & 1215.556 \\
& 1215.559 & 1215.623 & 1215.711 & 1221.556 \ 1224.572 & 1224.658 & 1225.665 & 1226.620 \\
& 1232.513 & 1232.604 & 1243.521 & 1243.577 \ 1243.601 & 1245.515 & 1245.567 & 1245.623 \\
& 1246.521 & 1246.573 & 1246.628 & 1263.532 &&&&\\
\hline
\bfseries{1999-2000}& 1513.622 & 1514.617 & 1518.647 & 1520.662 \ 1521.631 & 1522.651 & 1525.632 & 1529.649 \\
& 1529.707 & 1529.803 & 1530.638 & 1532.645 \ 1549.557 & 1549.629 & 1558.593 & 1558.690 \\
& 1563.521 & 1565.521 & 1570.554 & 1571.517 \ 1572.529 & 1572.612 & 1577.560 & 1580.506 \\
& 1580.568 & 1582.526 & 1583.542 & 1583.647 \ 1584.530 & 1586.554 & 1590.543 & 1590.606 \\
& 1592.512 & 1595.505 & 1596.524 & 1597.583 \ 1598.514 & 1604.522 & 1604.599 & 1607.529 \\
& 1608.524 & 1610.573 & 1612.589 & 1617.546 \ 1618.544 & 1618.576 & 1627.581 & 1630.572 \\
\hline
\bfseries{2000-2001}& 1837.689 & 1838.749 & 1840.651 & 1841.629 \ 1842.762 & 1843.693 & 1850.693 & 1866.816 \\
& 1870.685 & 1884.560 & 1886.506 & 1886.609 \ 1891.671 & 1897.545 & 1899.538 & 1913.551 \\
& 1913.692 & 1914.521 & 1921.524 & 1921.735 \ 1922.538 & 1922.626 & 1923.551 & 1923.698 \\
\hline
\bfseries{2001-2002}& 2220.676 & 2221.686 & 2224.758 & 2225.762 \ 2226.731 & 2227.621 & 2227.753 & 2230.640 \\
& 2231.609 & 2231.696 & 2231.838 & 2232.681 \ 2240.618 & 2253.610 & 2255.560 & 2259.588 \\
& 2300.629 & 2300.741 & 2301.594 & 2301.686 \ 2302.596 & 2302.690 & 2304.639 & 2309.548 \\
& 2309.701 & 2310.584 & 2314.596 & 2315.564 \ 2315.635 & 2317.588 & 2320.512 & 2322.564 \\
& 2322.617 & 2324.535 & 2324.604 & 2327.530 \ 2329.557 & 2329.628 & 2330.579 & 2334.553 \\
& 2335.586 & 2335.643 & 2341.592 & 2362.512 &&&&\\ 
\hline
\bfseries{2002-2003}& 2549.770 & 2553.815 & 2555.774 & 2562.744 \ 2569.698 & 2576.730 & 2587.696 & 2602.700 \\
& 2603.705 & 2612.719 & 2617.693 & 2617.782 \ 2621.631 & 2650.640 & 2658.536 & 2659.519 \\
& 2660.537 & 2662.566 & 2676.539 & 2676.627 \ 2683.542 & 2684.526 & 2684.632 & 2691.576 \\
& 2696.597 & 2697.622 & 2702.596 & 2709.596 \ 2714.595 & 2722.580 & 2723.567 & \\
\enddata
\end{deluxetable}
\clearpage

\begin{deluxetable}{|rrrrcrrrcl|}
\tablenum{2}
\tabletypesize{\scriptsize}
\tablecolumns{10}
\tablewidth{0pt}
\tablecaption{Variability Properties.\label{tab2}}
\tablehead{
\colhead{\bfseries{ HMW}}&\colhead{\bfseries{ LRLL}}&\colhead{\bfseries{R.A.}}&\colhead{\bfseries{Dec.}}&\colhead{\bfseries{R-I}}&\colhead{\bfseries{$\langle$I$\rangle$}}&\colhead{\bfseries{$\sigma_{var}$}}&\colhead{\bfseries{Range}}&\colhead{\bfseries{Period}}&\colhead{\bfseries{Notes}}\\
\colhead{}&\colhead{}&\colhead{J2000.0}&\colhead{J2000.0}&\colhead{Mag.}&\colhead{Mag.}&\colhead{Mag.}&\colhead{Mag.}&\colhead{days}&\colhead{}}
\startdata  
  1 & 1 & 3 44 34.19 & 32 09 46.0 & 0.52 & 7.67 & 0.076 & 0.455 & & a \\
  2 & 2 & 3 44 35.35 & 32 10 04.2 & 0.69 & 8.99 & 0.010 & 0.062 & & \\
  3 & 19 & 3 44 30.81 & 32 09 55.4 & 0.59 & 10.69 & 0.004 & 0.030 & & \\
  4 & 4 & 3 44 31.18 & 32 06 21.7 & 0.68 & 9.36 & 0.007 & 0.047 & & b \\
  5 & 8 & 3 44 09.12 & 32 07 09.0 & 0.38 & 9.35 & 0.020 & 0.105 & & \\
  6 & 7 & 3 44 08.44 & 32 07 16.2 & 0.36 & 9.30 & 0.018 & 0.095 & & \\
  7 & 28 & 3 44 21.04 & 32 07 38.3 & 0.47 & 10.67 & 0.006 & 0.033 & & \\
  8 & 30 & 3 44 19.11 & 32 09 30.8 & 0.53 & 10.86 & 0.005 & 0.027 & & \\
  9 & 38 & 3 44 23.97 & 32 10 59.5 & 0.83 & 11.86 & 0.019 & 0.100 & & \\ 
 10 & 10 & 3 44 24.65 & 32 10 14.7 & 0.71 & 10.29 & 0.005 & 0.030 & & \\
 11 & 6 & 3 44 36.94 & 32 06 45.1 & 1.03 & 10.80 & 0.017 & 0.094 & 1.687 & c,g \\
 12 & 29 & 3 44 31.52 & 32 08 44.6 & 0.83 & 12.15 & 0.032 & 0.181 & 2.237& c \\ 
 13 & 12 & 3 44 31.99 & 32 11 43.5 & 1.15 & 12.01 & 0.011 & 0.048 & & g \\
 14 & 9 & 3 44 39.16 & 32 09 17.8 & 1.39 & 12.12 & 0.024 & 0.109 & 2.539 & c,g \\ 
 15 & 35 & 3 44 39.23 & 32 07 35.1 & 1.40 & 13.15 & 0.212 & 0.715 & var. & d \\
 16 & 36 & 3 44 38.45 & 32 07 35.3 & 1.30 & 13.17 & 0.085 & 0.330 & 5.221 & \\ 
 17 & 77 & 3 44 43.43 & 32 08 17.0 & 0.96 & 12.79 & 0.008 & 0.039 & & \\
 18 & 59 & 3 44 40.11 & 32 11 33.7 & 1.26 & 13.49 & 0.034 & 0.170 & var. & d \\
 19 & & 3 44 25.56 & 32 12 29.5 & 1.15 & 13.50 & 0.053 & 0.247 & 8.363 & c \\ 
 20 & 5 & 3 44 26.01 & 32 04 30.1 & 1.35 & 12.42 & 0.103 & 0.754 & var. & d \\ 
 21 & & 3 44 27.00 & 32 04 43.4 & 1.27 & 13.61 & 0.025 & 0.115 & & \\
 22 & & 3 44 44.71 & 32 04 02.2 & 1.52 & 13.17 & 0.047 & 0.287 & var. & d \\
 23 & 37 & 3 44 37.96 & 32 03 29.4 & 1.18 & 13.39 & 0.342 & 1.978 & var. & d \\ 
 24 & & 3 44 04.97 & 32 09 53.4 & 1.07 & 13.04 & 0.024 & 0.124 & & \\
 25 & & 3 44 05.74 & 32 12 28.4 & 1.05 & 13.84 & 0.014 & 0.071 & & \\
 26 & & 3 44 16.40 & 32 09 54.8 & 0.89 & 12.65 & 0.048 & 0.230 & 3.021 & c \\
 27 & 16 & 3 44 32.74 & 32 08 37.0 & 0.85 & 11.62 & 0.017 & 0.097 & 2.656 & c \\ 
 28 & 33 & 3 44 32.58 & 32 08 42.0 & 1.62 & 12.51 & 0.481 & 1.462 & & e \\ 
 29 & 32 & 3 44 37.88 & 32 08 03.7 & 1.64 & 14.04 & 0.067 & 0.377 & var. & d \\
 30 & 58 & 3 44 38.54 & 32 08 00.2 & 1.63 & 14.10 & 0.068 & 0.490 & 7.570 & c \\
 31 & 62 & 3 44 26.61 & 32 03 58.0 & 1.82 & 14.03 & 0.048 & 0.204 & 3.087 & c \\ 
 32 & & 3 44 24.55 & 32 03 56.9 & 1.20 & 15.16 & 0.086 & 0.377 & 4.942 & c \\
 33 & 120 & 3 44 22.96 & 32 11 56.7 & 1.64 & 14.40 & 0.023 & 0.138 & & \\ 
 34 & & 3 44 22.30 & 32 12 00.2 & 1.27 & 14.52 & 0.087 & 0.501 & var. & d \\
 35 & & 3 44 21.28 & 32 11 55.9 & 1.87 & 14.66 & 0.025 & 0.144 & & \\
 36 & & 3 44 17.89 & 32 12 19.8 & 1.44 & 13.91 & 0.027 & 0.137 & & \\
 37 & 95 & 3 44 21.90 & 32 12 11.0 & 1.64 & 14.32 & 0.021 & 0.106 & & \\
 38 & & 3 44 21.75 & 32 12 30.9 & (2.00) & 14.96 & 0.024 & 0.155 & & f \\ 
 39 & & 3 44 23.65 & 32 06 46.2 & 1.74 & 14.12 & 0.029 & 0.251 & 9.667 & c,g \\
 40 & 125 & 3 44 21.63 & 32 06 24.5 & 1.64 & 14.49 & 0.036 & 0.175 & 8.363 & c \\ 
 41 & & 3 44 28.46 & 32 07 22.1 & 1.14 & 13.32 & 0.075 & 0.275 & 6.997 & c \\
 42 & 86 & 3 44 27.86 & 32 07 31.2 & 1.73 & 14.05 & 0.015 & 0.097 & 6.532 & c,g \\
 43 & 71 & 3 44 32.56 & 32 08 55.4 & 1.91 & 14.31 & 0.020 & 0.144 & & \\ 
 44 & 65 & 3 44 33.97 & 32 08 53.7 & 1.20 & 13.54 & 0.054 & 0.235 & 16.40 & c \\
 45 & 23 & 3 44 38.71 & 32 08 41.6 & 1.72 & 13.80 & 0.031 & 0.130 & 2.405 & c \\ 
 46 & 108 & 3 44 38.69 & 32 08 56.3 & 1.81 & 14.39 & 0.035 & 0.319 & & \\
 47 & 83 & 3 44 37.41 & 32 09 00.5 & 1.64 & 14.77 & 0.146 & 1.066 & 8.378 & d,i \\
 48 & 113 & 3 44 37.19 & 32 09 15.7 & 1.22 & 14.25 & 0.030 & 0.155 & & \\
 49 & & 3 44 37.40 & 32 06 11.3 & 1.29 & 13.77 & 0.018 & 0.103 & 6.221 & \\
 50 & & 3 44 34.86 & 32 06 33.3 & 1.18 & 13.27 & 0.048 & 0.494 & 5.483 & c \\ 
51 & & 3 44 42.61 & 32 06 19.0 & 1.13 & 13.89 & 0.020 & 0.102 & 11.51 & c \\
 52 & & 3 44 11.22 & 32 06 11.7 & 1.59 & 14.21 & 0.039 & 0.175 & 10.77 & c \\
53 & & 3 44 38.37 & 32 12 59.2 & (1.77) & 14.55 & 0.036 & 0.198 & 13.48 & c,h \\ 
 54 & 74 & 3 44 34.25 & 32 10 49.2 & 1.98 & 14.38 & 0.021 & 0.142 & & \\
 55 & & 3 44 49.61 & 32 09 11.5 & 1.09 & 14.98 & 0.022 & 0.164 & & \\ 
 56 & 75 & 3 44 43.77 & 32 10 29.9 & 1.59 & 14.36 & 0.229 & 1.358 & var. & d \\
 57 & & 3 44 18.14 & 32 04 56.7 & 1.95 & 15.24 & 0.062 & 0.478 & & \\
 58 & & 3 44 14.10 & 32 10 27.9 & 1.49 & 14.69 & 0.022 & 0.110 & & \\
 59 & 41 & 3 44 21.59 & 32 10 37.2 & 1.63 & 15.10 & 0.251 & 1.177 & var. & d \\
 60 & & 3 44 21.54 & 32 10 17.0 & 1.65 & 14.54 & 0.089 & 0.481 & 7.092 & c \\ 
 61 & 128 & 3 44 20.15 & 32 08 56.2 & 1.92 & 14.80 & 0.022 & 0.154 & & \\ 
 62 & & 3 44 04.07 & 32 07 16.7 & 1.64 & 14.94 & 0.035 & 0.224 & & \\
 63 & & 3 44 06.77 & 32 04 40.6 & 1.70 & 15.06 & 0.028 & 0.190 & & \\
 64 & & 3 44 21.23 & 32 05 02.2 & 1.74 & 15.06 & 0.053 & 0.283 & & \\
 65 & & 3 44 22.26 & 32 05 42.4 & 1.48 & 14.95 & 0.196 & 0.979 & var. & d \\
 66 & & 3 44 35.69 & 32 04 52.3 & 2.42 & 16.03 & 0.061 & 0.480 & & \\
 67 & & 3 44 38.59 & 32 05 06.0 & 2.15 & 14.80 & 0.019 & 0.164 & & \\
 68 & & 3 44 31.65 & 32 06 53.3 & 2.55 & 16.10 & 0.048 & 0.435 & & \\
 69 & & 3 44 37.77 & 32 12 17.7 & (1.88) & 15.39 & 0.078 & 0.355 & 3.256 & e,h \\
 70 & & 3 44 11.19 & 32 08 15.8 & 2.62 & 15.13 & 0.036 & 0.248 & & \\
 71 & & 3 44 6.75 & 32 07 53.7 & 2.24 & 15.29 & 0.040 & 0.305 & & \\
 72 & & 3 44 6.10 & 32 07 06.6 & 2.50 & 15.89 & 0.052 & 0.527 & & g \\
 73 & & 3 44 19.22 & 32 07 34.4 & 1.80 & 14.88 & 0.257 & 1.271 & 7.582 & c,e \\
 74 & & 3 44 18.24 & 32 07 32.2 & 2.27 & 16.04 & 0.058 & 0.662 & & \\
 75 & 40 & 3 44 29.71 & 32 10 39.4 & 1.44 & 14.18 & 0.054 & 0.337 & var. & d,g \\
 76 & 88 & 3 44 32.75 & 32 09 15.3 & (2.00) & 14.62 & 0.043 & 0.240 & 5.286 & c,f \\
 77 & 151 & 3 44 34.81 & 32 11 17.5 & 2.08 & 14.91 & 0.022 & 0.145 & & \\
 78 & 45 & 3 44 24.27 & 32 10 19.0 & 1.08 & 12.64 & 0.201 & 1.221 & & e \\
 79 & & 3 44 25.25 & 32 10 12.6 & 2.37 & 14.48 & 1.052 & 4.816 & & e \\
 80 & 96 & 3 44 34.83 & 32 09 52.6 & 2.11 & 13.63 & 1.724 & 7.644 & & e \\
 81 & 91 & 3 44 39.19 & 32 09 44.4 & 1.86 & 14.69 & 0.059 & 0.445 & 3.993 & c \\
 82 & 145 & 3 44 41.29 & 32 10 24.7 & 2.22 & 14.75 & 0.021 & 0.190 & & \\
 83 & & 3 44 43.04 & 32 10 14.6 & 2.55 & 16.09 & 0.059 & 0.626 & & \\
 84 & 42 & 3 44 42.05 & 32 09 00.2 & 1.98 & 14.31 & 0.110 & 0.549 & & e \\
 85 & & 3 44 42.07 & 32 09 00.7 & 1.87 & 14.34 & 0.114 & 0.550 & & e \\
 86 & & 3 44 31.40 & 32 11 28.9 & (2.00) & 15.82 & 0.043 & 0.318 & & f \\
 87 & 193 & 3 44 37.99 & 32 11 36.6 & 2.20 & 15.56 & 0.092 & 0.644 & & \\ 
 88 & & 3 44 37.38 & 32 12 23.6 & (1.96) & 15.85 & 0.106 & 0.955 & & e,h \\ 
 89 & & 3 44 40.78 & 32 13 06.1 & (2.00) & 15.97 & 0.054 & 0.357 & & f \\
 90 & & 3 44 41.43 & 32 13 09.2 & (2.00) & 15.72 & 0.041 & 0.257 & & f \\ 
 91 & & 3 44 48.82 & 32 13 21.3 & (2.00) & 14.94 & 0.035 & 0.190 & & f \\
 92 & & 3 44 41.72 & 32 12 01.8 & (2.00) & 16.13 & 0.067 & 0.430 & & f \\
 93 & 404 & 3 44 44.42 & 32 10 05.1 & 1.65 & 16.67 & 0.086 & 0.576 & & \\
 94 & 171 & 3 44 44.83 & 32 11 05.1 & 1.94 & 15.31 & 0.030 & 0.253 & & \\
 95 & 360 & 3 44 43.69 & 32 10 47.4 & 2.09 & 16.46 & 0.073 & 0.612 & & \\
 96 & 277 & 3 44 39.43 & 32 10 07.4 & 2.22 & 16.10 & 0.047 & 0.376 & & \\
 97 & 167 & 3 44 41.15 & 32 10 09.6 & 2.47 & 16.66 & 0.139 & 0.892 & & \\
 98 & 166 & 3 44 42.58 & 32 10 02.2 & 2.83 & 16.84 & 0.091 & 0.960 & & e \\
 99 & 221 & 3 44 40.25 & 32 09 32.4 & 2.41 & 16.60 & 0.157 & 2.136 & & g \\
100 & 442 & 3 44 40.67 & 32 09 40.6 & 1.32 & 17.43 & 0.182 & 1.658 & & \\
101 & & 3 44 18.17 & 32 09 58.9 & 2.54 & 15.78 & 0.129 & 0.749 & var. & d \\
102 & & 3 44 19.19 & 32 09 59.8 & (2.00) & 18.92 & 0.512 & 2.350 & & f \\
103 & & 3 44 18.53 & 32 09 46.3 & (2.00) & 18.10 & 0.830 & 2.798 & & f \\
104 & 294 & 3 44 24.56 & 32 10 03.0 & 2.09 & 16.05 & 0.147 & 1.106 & & e \\
105 & 237 & 3 44 23.55 & 32 09 33.4 & 2.30 & 15.70 & 0.041 & 0.385 & & \\
106 & 90 & 3 44 33.31 & 32 09 39.6 & 1.83 & 14.07 & 0.255 & 2.430 & & e \\
107 & 165 & 3 44 35.46 & 32 08 56.0 & 2.60 & 15.96 & 0.054 & 0.328 & & e \\
108 & 149 & 3 44 36.98 & 32 08 33.8 & 2.41 & 15.58 & 0.062 & 0.597 & & \\
109 & 153 & 3 44 42.76 & 32 08 33.4 & 2.25 & 15.78 & 0.067 & 0.458 & & \\
110 & 103 & 3 44 44.58 & 32 08 12.1 & 1.82 & 15.35 & 0.163 & 0.910 & var. & d \\
111 & 52 & 3 44 43.52 & 32 07 42.3 & 1.85 & 14.82 & 0.029 & 0.187 & & e \\
112 & & 3 44 34.43 & 32 06 24.6 & 2.59 & 15.91 & 0.054 & 0.354 & & \\
113 & & 3 44 49.98 & 32 03 45.2 & (2.00) & 14.97 & 0.052 & 0.660 & & f \\
114 & & 3 44 49.78 & 32 03 33.7 & (2.00) & 15.17 & 0.034 & 0.206 & & f \\ 
115 & & 3 44 34.34 & 32 04 21.5 & 1.62 & 16.47 & 0.074 & 0.595 & & \\ 
116 & & 3 44 32.79 & 32 04 12.9 & 1.24 & 16.09 & 0.051 & 0.474 & & \\
117 & & 3 44 38.84 & 32 06 35.9 & 2.62 & 16.54 & 0.070 & 0.538 & & \\
118 & & 3 44 41.22 & 32 06 26.6 & 2.50 & 16.73 & 0.089 & 0.805 & & \\
119 & 141 & 3 44 30.53 & 32 06 29.3 & 1.75 & 14.68 & 0.140 & 1.016 & & e \\
120 & & 3 44 29.68 & 32 05 51.9 & 1.56 & 15.97 & 0.047 & 0.386 & & \\
121 & & 3 44 31.02 & 32 05 45.6 & 2.54 & 16.64 & 0.074 & 0.666 & & \\
122 & & 3 44 27.27 & 32 07 17.4 & 2.24 & 16.60 & 0.086 & 0.685 & & \\
123 & & 3 44 05.25 & 32 08 02.1 & 1.40 & 16.37 & 0.082 & 0.956 & & \\
124 & & 3 44 08.03 & 32 06 56.0 & 1.77 & 16.18 & 0.111 & 0.911 & & \\
125 & & 3 44 07.67 & 32 05 04.7 & 2.45 & 16.62 & 0.074 & 0.510 & & \\
126 & & 3 44 07.46 & 32 04 08.5 & 1.96 & 15.53 & 0.036 & 0.330 & & \\
127 & & 3 44 10.09 & 32 04 04.2 & (2.00) & 15.78 & 0.060 & 0.533 & & f \\
128 & & 3 44 32.41 & 32 05 11.4 & (2.00) & 19.07 & 0.481 & 1.945 & & e,f \\
129 & & 3 44 32.36 & 32 05 11.3 & (2.00) & 19.00 & 0.446 & 1.935 & & e,f \\
130 & & 3 44 21.20 & 32 12 36.8 & (2.00) & 15.69 & 0.151 & 1.710 & & e,f \\
131 & & 3 44 20.90 & 32 12 37.1 & (2.00) & 15.55 & 0.088 & 0.448 & & e,f \\ 
132 & & 3 44 18.57 & 32 12 52.7 & (2.00) & 16.56 & 0.345 & 2.005 & var. & d,f \\
133 & 60 & 3 44 25.55 & 32 11 30.1 & 1.90 & 14.78 & 0.101 & 0.485 & var. & d \\
134 & & 3 44 19.99 & 32 06 45.1 & (2.00) & 15.72 & 0.070 & 0.570 & 8.687 & f \\
135 & & 3 44 25.53 & 32 06 16.7 & 2.24 & 15.86 & 0.068 & 0.354 & & \\
136 & & 3 44 23.59 & 32 07 11.3 & 1.36 & 16.60 & 0.092 & 0.789 & & g \\
137 & & 3 44 34.04 & 32 06 56.8 & 2.80 & 16.87 & 0.120 & 0.891 & & \\
138 & & 3 44 04.20 & 32 09 38.1 & 0.95 & 15.02 & 0.029 & 0.275 & & \\
139 & & 3 44 09.97 & 32 09 41.3 & 1.31 & 16.15 & 0.060 & 0.488 & & \\
140 & & 3 44 25.58 & 32 10 44.8 & (2.00) & 19.21 & 0.441 & 1.040 & & e \\
141 & 194 & 3 44 27.22 & 32 10 36.7 & 2.38 & 16.22 & 0.206 & 1.264 & var. & d \\
142 & 230 & 3 44 35.51 & 32 08 04.1 & 2.54 & 16.15 & 0.059 & 0.485 & & \\
143 & & 3 44 35.03 & 32 07 36.6 & 1.23 & 12.72 & 0.039 & 0.226 & & e \\
144 & & 3 44 35.17 & 32 07 36.4 & (2.00) & 12.72 & 0.039 & 0.226 & & e,f \\
145 & & 3 44 20.26 & 32 05 43.5 & 2.57 & 16.91 & 0.110 & 0.963 & & \\
146 & & 3 44 19.14 & 32 05 59.5 & 2.63 & 16.83 & 0.098 & 0.630 & & \\
147 & 169 & 3 44 17.73 & 32 04 47.4 & 2.57 & 15.69 & 0.048 & 0.413 & & \\
148 & & 3 44 16.48 & 32 05 32.5 & 1.95 & 16.41 & 0.080 & 0.450 & & \\
149 & & 3 44 46.93 & 32 05 36.5 & 1.13 & 15.89 & 0.048 & 0.432 & & \\
150 & & 3 44 48.37 & 32 06 27.9 & (2.00) & 18.76 & 0.505 & 1.682 & & f \\
151 & & 3 44 50.78 & 32 07 00.5 & (2.00) & 18.96 & 0.387 & 1.488 & & f \\
\enddata
\tablenotetext{a}{This is a double star with a magnitude difference of 0.2.}
\tablenotetext{b}{\citet{ri02} have reported the detection of Delta Scuti-like pulsations in this star.}
\tablenotetext{c}{Different periods are detected in different seasons, and the period given here is an average over all five seasons.}
\tablenotetext{d}{Stars which vary by more than 3$\sigma$ are marked ``var.''}
\tablenotetext{e}{Photometry may be contaminated by a nearby star.  See table 3 of \citet{hmw}.}
\tablenotetext{f}{For stars with no R-I or spectral type known, an R-I of 2.00 is assumed (see text).}
\tablenotetext{g}{Several stars were detected by \citet{dbs} as binaries with separations of less than 3$\arcsec$.}
\tablenotetext{h}{The given R-I value is an average of R-I values of stars with similar magnitudes and spectral types (see text).}
\tablenotetext{i}{Star 47 is periodic is the 1998-99 season, but in subsequent seasons undergoes large-amplitude irregular variations (see section 4).}
\end{deluxetable}
\clearpage

\begin{deluxetable}{cccccccc}
\tablenum{3}
\tabletypesize{\scriptsize}
\tablecaption{Cousins I Photometry of Star 15 \label{table3}}
\tablewidth{0pt}
\tablehead{\colhead{JD-2450000} & \colhead{I (mag)}   & 
\colhead{JD-2450000} & \colhead{I (mag)}   & 
\colhead{JD-2450000} & \colhead{I (mag)}   & 
\colhead{JD-2450000} & \colhead{I (mag)}}\startdata
1157.687 & 12.96 & 1157.695 & 12.96 & 1162.581 & 12.96 & 1162.680 & 12.97 \\
1163.644 & 12.97 & 1163.658 & 12.96 & 1163.666 & 12.97 & 1164.598 & 12.97 \\
1166.505 & 12.96 & 1166.513 & 12.96 & 1170.799 & 12.96 & 1184.464 & 12.96 \\
1210.569 & 12.94 & 1215.556 & 12.95 & 1215.623 & 12.96 & 1243.521 & 12.97 \\ 
1263.532 & 12.97 & 1243.521 & 12.97 & 1166.581 & 12.96 & 1166.589 & 12.97 \\
1166.610 & 12.95 & 1166.618 & 12.96 & 1170.748 & 12.95 & 1170.790 & 12.96 \\
1170.799 & 12.96 & 1178.561 & 12.98 & 1178.569 & 12.98 & 1178.603 & 12.98 \\
1178.611 & 12.96 & 1178.702 & 12.96 & 1178.710 & 12.95 & 1183.496 & 12.96 \\
1183.522 & 12.97 & 1183.627 & 12.97 & 1183.669 & 12.97 & 1183.727 & 12.97 \\
1184.464 & 12.96 & 1184.657 & 12.97 & 1184.782 & 12.98 & 1186.525 & 12.97 \\
1186.599 & 12.97 & 1186.745 & 12.97 & 1190.588 & 12.97 & 1196.564 & 12.97 \\
1198.515 & 12.97 & 1198.683 & 12.97 & 1199.484 & 12.96 & 1199.690 & 12.98 \\
1209.515 & 12.96 & 1209.534 & 12.96 & 1209.592 & 12.96 & 1209.686 & 12.97 \\
1210.569 & 12.97 & 1210.649 & 12.97 & 1213.526 & 12.96 & 1215.556 & 12.97 \\
1215.559 & 12.97 & 1215.623 & 12.96 & 1215.711 & 12.97 & 1221.556 & 12.98 \\
1224.572 & 12.97 & 1224.658 & 12.96 & 1225.665 & 12.98 & 1226.620 & 12.96 \\
1232.513 & 12.97 & 1232.604 & 12.97 & 1243.521 & 12.97 & 1243.577 & 12.96 \\
1243.601 & 12.97 & 1245.515 & 12.98 & 1245.567 & 12.97 & 1245.623 & 12.97 \\
1246.521 & 12.97 & 1246.573 & 12.98 & 1246.628 & 12.98 & 1263.532 & 12.98 \\
1513.622 & 13.16 & 1514.617 & 13.16 & 1518.647 & 13.17 & 1520.662 & 13.18 \\
1521.631 & 13.17 & 1522.651 & 13.18 & 1525.632 & 13.19 & 1529.649 & 13.19 \\
1529.707 & 13.19 & 1529.803 & 13.18 & 1530.638 & 13.20 & 1532.645 & 13.19 \\
1549.557 & 13.22 & 1549.629 & 13.22 & 1558.593 & 13.24 & 1558.690 & 13.23 \\
1563.521 & 13.25 & 1565.521 & 13.25 & 1570.554 & 13.25 & 1571.517 & 13.24 \\
1572.529 & 13.26 & 1572.612 & 13.26 & 1577.560 & 13.25 & 1580.506 & 13.26 \\
1580.568 & 13.27 & 1582.526 & 13.27 & 1583.542 & 13.27 & 1583.647 & 13.27 \\
1584.530 & 13.26 & 1586.554 & 13.28 & 1590.543 & 13.28 & 1590.606 & 13.28 \\
1592.512 & 13.28 & 1595.505 & 13.30 & 1596.524 & 13.29 & 1597.583 & 13.30 \\
1598.514 & 13.29 & 1604.522 & 13.29 & 1604.599 & 13.29 & 1607.529 & 13.29 \\
1608.524 & 13.31 & 1610.573 & 13.31 & 1612.589 & 13.33 & 1617.546 & 13.32 \\
1618.544 & 13.33 & 1618.576 & 13.32 & 1627.581 & 13.33 & 1630.572 & 13.33 \\
1837.689 & 13.64 & 1838.749 & 13.65 & 1840.651 & 13.64 & 1841.629 & 13.65 \\
1842.762 & 13.64 & 1843.693 & 13.65 & 1850.693 & 13.65 & 1866.816 & 13.64 \\
1870.685 & 13.65 & 1884.560 & 13.63 & 1886.506 & 13.63 & 1886.609 & 13.64 \\
1891.671 & 13.62 & 1897.545 & 13.65 & 1899.538 & 13.65 & 1913.551 & 13.65 \\
1913.692 & 13.64 & 1914.521 & 13.65 & 1921.524 & 13.65 & 1921.735 & 13.63 \\
1922.538 & 13.63 & 1922.626 & 13.64 & 1923.551 & 13.65 & 1923.698 & 13.66 \\
2220.676 & 13.25 & 2221.686 & 13.24 & 2224.758 & 13.24 & 2225.762 & 13.25 \\
2226.731 & 13.25 & 2227.621 & 13.26 & 2227.753 & 13.24 & 2230.640 & 13.25 \\
2231.609 & 13.26 & 2231.696 & 13.25 & 2231.838 & 13.24 & 2232.681 & 13.24 \\
2240.618 & 13.24 & 2253.610 & 13.22 & 2255.560 & 13.22 & 2259.588 & 13.21 \\
2300.629 & 13.18 & 2300.741 & 13.17 & 2301.594 & 13.18 & 2301.686 & 13.17 \\
2302.596 & 13.17 & 2302.690 & 13.17 & 2304.639 & 13.18 & 2309.548 & 13.17 \\
2309.701 & 13.17 & 2310.584 & 13.16 & 2314.596 & 13.14 & 2315.564 & 13.15 \\
2315.635 & 13.16 & 2317.588 & 13.14 & 2320.512 & 13.15 & 2322.564 & 13.15 \\
2322.617 & 13.14 & 2324.535 & 13.14 & 2324.604 & 13.15 & 2327.530 & 13.14 \\ 
2329.557 & 13.13 & 2329.628 & 13.14 & 2330.579 & 13.13 & 2334.553 & 13.12 \\
2335.586 & 13.11 & 2335.643 & 13.11 & 2341.592 & 13.09 & 2362.512 & 13.08 \\
2549.770 & 13.02 & 2553.815 & 13.02 & 2555.774 & 13.02 & 2562.744 & 13.02 \\
2569.698 & 13.02 & 2576.730 & 13.00 & 2587.696 & 13.01 & 2602.700 & 12.98 \\
2603.705 & 12.99 & 2612.719 & 12.98 & 2617.693 & 12.98 & 2617.782 & 12.98 \\
2621.631 & 12.98 & 2650.640 & 12.98 & 2658.536 & 12.97 & 2659.519 & 12.98 \\
2660.537 & 12.97 & 2662.566 & 12.97 & 2676.539 & 12.98 & 2676.627 & 12.98 \\ 
2683.542 & 12.98 & 2684.526 & 12.98 & 2684.632 & 12.98 & 2691.576 & 12.98 \\
2696.597 & 12.99 & 2697.622 & 12.99 & 2702.596 & 12.99 & 2709.596 & 12.98 \\ 
2714.595 & 12.99 & 2722.580 & 12.98 & 2723.567 & 12.99 & & \\
\enddata
\end{deluxetable}

\end{document}